\documentclass{article}

\usepackage{arxiv}

\usepackage[utf8]{inputenc}
\usepackage[T1]{fontenc}
\usepackage{microtype}
\usepackage{url}
\usepackage{amsmath}
\usepackage{amssymb}
\usepackage{amsfonts}
\usepackage{bm}
\usepackage{nicefrac}
\usepackage{graphicx}
\usepackage{float}
\usepackage{subcaption}
\usepackage{booktabs}
\usepackage{dcolumn}
\usepackage{xcolor}
\usepackage[authoryear,round]{natbib}
\usepackage{hyperref}
\usepackage{doi}

\title{NeRF-Inspired Neural Fields for 3D Super-resolution of Reacting-Flows}

\author{
\href{https://orcid.org/0000-0003-0228-8848}{Dakshaka Vishwajith Amarathunga}
\quad
\href{https://orcid.org/0009-0006-2001-337X}{Shang-Ru Li}
\quad
\href{https://orcid.org/0000-0002-0146-7994}{Tarek Echekki}\thanks{Corresponding author: \texttt{techekk@ncsu.edu}} \\
Department of Mechanical and Aerospace Engineering \\
North Carolina State University \\
Campus Box 7910, Raleigh, NC 27695, USA
}

\renewcommand{\shorttitle}{Neural Field Super-Resolution of Reacting Flows}

\date{}

\hypersetup{
    pdftitle={NeRF-Inspired Neural Fields for 3D Super-resolution of Reacting-Flows},
    pdfauthor={Dakshaka Vishwajith Amarathunga,
               Shang-Ru Li,
               Tarek Echekki},
    pdfsubject={Neural Field Super-Resolution of Reacting Flows},
    pdfkeywords={Turbulent Reacting Flows, Combustion,
                 Super-Resolution, Machine Learning,
                 Computational Methods}
}

\begin{document}
\maketitle

\begin{abstract}
The multiscale nature of turbulent reacting flows and the importance of fine-scale structures, including thin reaction layers, pose challenges for super-resolution. In this study, we investigate a 3D super-resolution framework inspired by Neural Radiance Fields (NeRF), called Neural Field Super-Resolution (NFSR), to reconstruct fine-resolution fields from coarsened reacting-flow fields. The proposed formulation adapts the continuous
coordinate-conditioned representation of NeRF to physical flow-field
reconstruction by conditioning spatial coordinates with local coarse-grid thermochemical
information, while eliminating radiance prediction and volume rendering. Principal
component analysis is used to provide a compact and low-dimensional representation of the thermochemical
state, and a hierarchical extension, H-NFSR, progressively reconstructs the solution
through intermediate resolution levels. The framework is evaluated using
three-dimensional direct numerical simulation data for a turbulent premixed hydrogen-air flame from the BLASTNet 3.0 dataset, with an overall spatial coarsening factor of eight in each coordinate direction. The results demonstrate consistently better prediction measures and a higher ability to reconstruct fine-scale information than tricubic interpolation on the unseen test realization. The hierarchical formulation reduces
visible coarse-cell imprinting. Training
with a mean absolute error loss provides modest improvements in the overall quantitative
performance compared with a mean squared error loss. In contrast, biased sampling mainly
benefits the most localized structures. Three-dimensional isosurface comparisons further
show that H-NFSR recovers much of the small and intermediate features of the flow, including surface wrinkling and local flame curvature effects. These
results highlight the potential of coordinate-conditioned neural fields as a flexible
framework for three-dimensional coarse-to-fine reconstruction of reacting-flow
simulations.
\end{abstract}


\maketitle

\section{Introduction}
\label{sec:introduction}

Turbulent reacting flows involve a wide range of interacting spatial and temporal scales.
Thin reaction zones, steep thermochemical gradients, and highly wrinkled flame structures
can require very fine spatial resolution for accurate numerical representation. Direct
numerical simulation (DNS)~\citep{Domingo2022} provides access to these fine-scale features, but its
computational cost becomes prohibitive for many practical reacting-flow configurations.
Consequently, large-eddy simulation (LES), quasi-DNS~\citep{komen2017quantification, zirwes2020qdns, abe2026qdns}, and other coarse-grid approaches
are commonly used to reduce the computational requirements. In these approaches,
however, part of the fine-scale spatial information is either modeled or not directly
represented on the computational grid. This has motivated the development of data-driven
approaches that seek to recover fine-resolution information from coarse grid flow
representations based on super-resolution.

A number of machine-learning approaches have been developed for flow-field
super-resolution. Convolutional neural networks (CNNs) have demonstrated the ability
to reconstruct fine-scale turbulent structures from low-resolution inputs
\citep{Fukami2019,Fukami2023}. In reacting flows, related reconstruction and
subgrid-scale modeling problems have been addressed using deconvolution- and
CNN-based approaches for quantities associated with scalar transport, reaction rates,
and filtered thermochemical fields
\citep{Nikolaou2018,Lapeyre2019,Seltz2019}. Generative Adversarial Networks (GANs)
and their super-resolution variants have also been applied to turbulent and reacting-flow
reconstruction, where adversarial training is used to improve the representation of
small-scale spatial structures
\citep{Goodfellow2020,Wang-ESRGAN,Bode2021,Grenga2023,Nista2023,Nista2024,Nista2025}.
More recently, Graph Neural Networks (GNNs) have been introduced for flow-field
super-resolution, particularly for problems involving irregular or unstructured meshes
\citep{Barwey2025,Dash2026}, in addition to techniques based on diffusion models~\citep{Huang2024, Chen2026}.

Despite their demonstrated capabilities, these approaches rely on different
architecture-specific representations of the spatial field. CNN-based methods generally
operate on structured spatial patches or volumes, GAN-based formulations introduce an
additional adversarial training procedure. At the same time, GNN-based approaches require an
explicit graph representation of the computational domain. Moreover, these approaches
typically represent the reconstructed solution on a prescribed discrete grid or graph,
rather than as a continuous spatial field. This can limit the direct evaluation of the
solution at arbitrary spatial coordinates and motivates the exploration of alternative
super-resolution formulations based on continuous field representations. In this context,
coordinate-conditioned neural representations provide a natural framework in which the
fine-resolution solution can be evaluated directly at prescribed spatial locations.

An extension of the Neural Implicit Flow (NIF), which was originally developed by~\citet{Pan2023}, was proposed by~\citet{Amarathunga-ESSCI} to reconstruct super-resolution fields from coarse simulation data of reacting flows. NIF provides a mesh-agnostic representation of spatio-temporal
flow fields by separating the spatial representation from temporal or parametric
information. In the NIF-based super-resolution framework, coarse-grid thermochemical
information is used to condition the reconstruction of fine-resolution reacting-flow
scalar fields. This formulation demonstrated the potential of coordinate-based neural
representations for continuous super-resolution of reacting-flow fields. However, the
framework relies on a hypernetwork architecture in which the parameters of the spatial
reconstruction network are generated dynamically from the conditioning information.
Although this provides a flexible representation, it also introduces additional
computational overhead. These disadvantages encourage the development of a more direct coordinate-conditioned formulation
in which coarse-grid flow information conditions the fine-resolution prediction without
dynamically generating the weights of the spatial network. Such a formulation is
particularly attractive for extending continuous coordinate-based super-resolution to
three-dimensional reacting flows.

The Neural Radiance Fields (NeRF) model introduced by
\citet{mildenhall2020nerf} provides a useful basis for such a formulation. NeRF
represents a three-dimensional scene as a continuous function of spatial coordinates and
employs positional encoding to improve the representation of high-frequency spatial
variation. Although originally developed for radiance-field reconstruction and novel-view
synthesis, the underlying coordinate-conditioned representation provides a useful
framework for physical fields characterized by strong localized spatial variations. This
motivates the present formulation of a NeRF-inspired neural-field framework for
three-dimensional super-resolution of reacting flows.

In the current study, a Neural Field Super-Resolution (NFSR) framework is introduced
to reconstruct three-dimensional fine-resolution thermochemical scalar fields from
coarse reacting flow data. The proposed framework provides a direct mapping between
coarse-grid information, spatial coordinates, and the corresponding fine-resolution
solution, while avoiding the dynamic parameter generation. A hierarchical extension is also examined to determine whether progressive coarse-to-fine reconstruction improves the recovery of localized structures and spatial continuity. In addition to the reconstruction architecture itself, the effects of training-data sampling and the optimization objective are investigated to determine how these modeling choices influence the reconstruction accuracy.

The training objective is of particular interest because the relative behavior of commonly
used loss functions can depend on the underlying super-resolution formulation.
\citet{lossstudyZhao} demonstrated that networks trained using an $L_1$ loss can provide
improved reconstruction quality compared with the commonly used $L_2$ loss for
image restoration and super-resolution problems, including improved preservation of
perceptually important structures. Their results also showed that the relative performance
of the loss functions can depend on the convergence behavior of the network rather than
only on the form of the error metric itself. More recently,
\citet{balla2025implicitaugmentationdistributionalsymmetry} adopted an MAE-based loss ($L_1$)
for three-dimensional turbulence super-resolution, motivated by previous observations
that MAE can better preserve reconstruction quality and reduce over-smoothing compared
with MSE ($L_2$) in image-restoration applications. These observations establish the need for an examination of whether similar behavior is observed in the present coordinate-conditioned NFSR framework.

The proposed framework is evaluated using three-dimensional DNS data for a turbulent
premixed hydrogen--air flame from the BLASTNet 3.0 dataset
\citep{blastnet3.0_2026}. A spatial coarsening factor of eight in each coordinate
direction is considered, resulting in a challenging reconstruction problem in which a
substantial portion of the fine-scale spatial information is absent from the coarse
representation. The study examines local and global reconstruction performance,
compares the learned models with tricubic interpolation, evaluates the effects of
hierarchical reconstruction, biased sampling, and loss function, and assesses the recovery
of the three-dimensional flow structure. The objective is to establish a simple and
flexible coordinate-conditioned framework for recovering fine-scale thermochemical
information from highly coarsened three-dimensional reacting-flow fields.

\section{Methodology}
\label{sec:Methodology}

\subsection{NeRF-inspired Neural-field Architecture}
\label{sec:nerf_architecture}

\begin{figure}[t]
        \centering
        \includegraphics[width=0.80\textwidth]{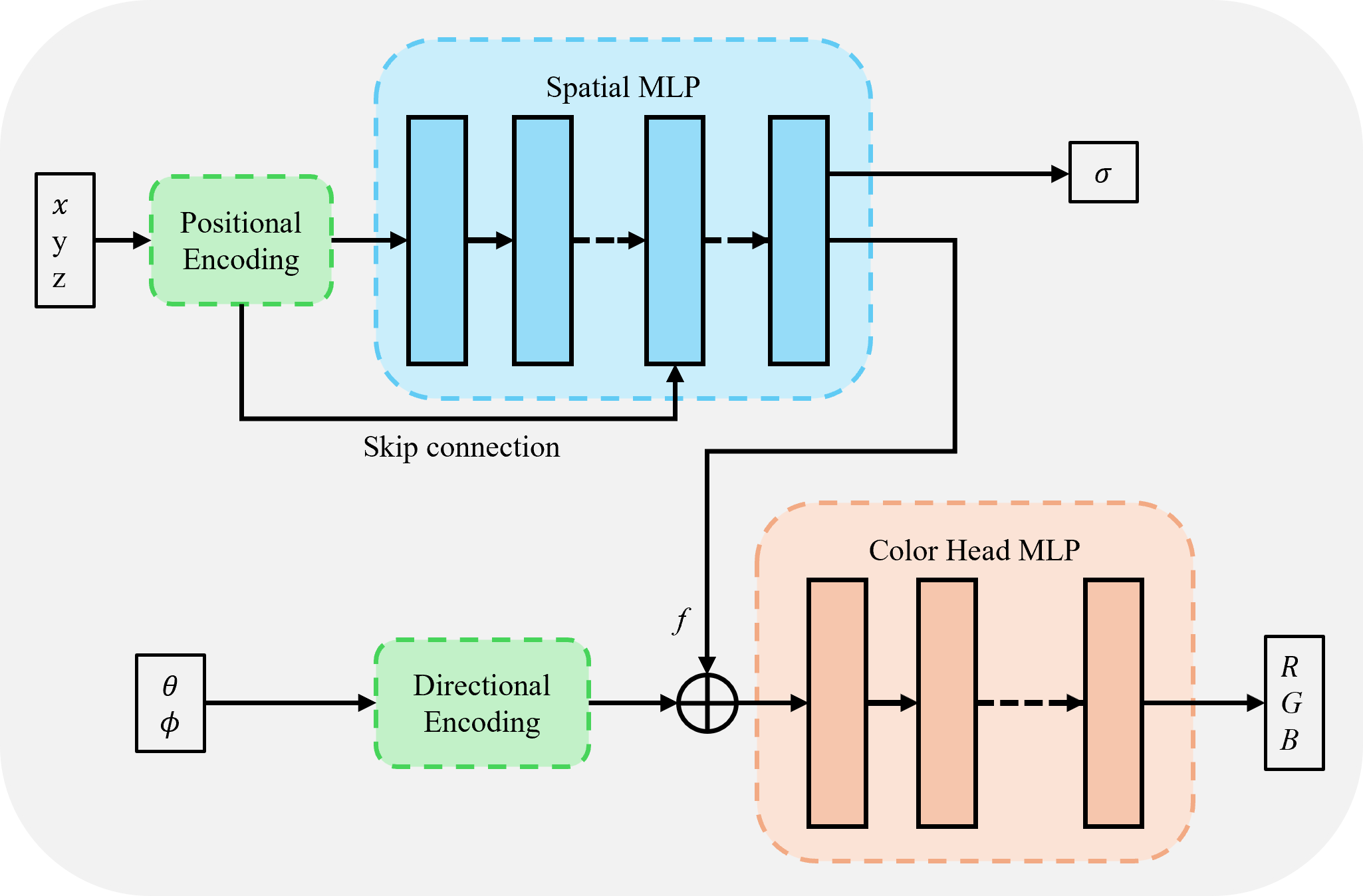}
        \caption{Conceptual schematic diagram of Neural Radiance Field (NeRF) architecture.}
        \label{fig:1}
\end{figure}

The proposed framework is based on an extension of the NeRF
architecture~\citep{mildenhall2020nerf}. Figure~\ref{fig:1} illustrates a
conceptual representation of the original NeRF architecture, which was originally developed
to represent a three-dimensional scene as a continuous function of the spatial coordinates
and the viewing direction. The network maps a spatial location,
$\boldsymbol{x}=(x,y,z)$, and a viewing direction, $\boldsymbol{d}$, to a volume density,
$\sigma$, and the corresponding emitted radiance, represented by 
components enclosed within the dashed lines. This mapping can be expressed as
\begin{equation}
F_{\boldsymbol{\Theta}}(\boldsymbol{x},\boldsymbol{d})
=
\left(\sigma,\boldsymbol{c}\right),
\qquad
\boldsymbol{c}=(R,G,B),
\label{eq:original_nerf}
\end{equation}
where $\boldsymbol{\Theta}$ denotes the trainable parameters of the network.

In the original NeRF architecture, the spatial coordinates are first transformed using a
positional encoding and passed through a fully connected spatial multilayer perceptron
(MLP), as shown in Fig.~\ref{fig:1}. The positional encoding maps the coordinates into a
higher-dimensional representation using sinusoidal functions and improves the ability of
the network to capture high-frequency spatial variations. A skip connection is also
incorporated within the spatial MLP by reintroducing the encoded coordinates at an
intermediate layer. The spatial MLP predicts the volume density, $\sigma$, and an
intermediate feature vector, $\boldsymbol{f}$. The feature vector is subsequently
concatenated with the encoded viewing direction and passed through a color-head MLP to
predict the view-dependent RGB radiance. The predicted density and radiance are evaluated
at multiple points along a camera ray and combined using a differentiable volume-rendering
procedure to construct the final image.

The present study modifies the NeRF architecture to address 1) the implementation of this architecture for super-resolution of scalar and velocity fields, 2) the multi-dimensional nature of chemical reacting flow states, and 3) the inherent challenge of capturing fine-scale structures from coarse-scale fields when such structures originate at the finest scales. 

When adapting the NeRF architecture to reacting-flow
super-resolution, the high dimensionality of the
thermochemical state must be addressed. The state may consist of temperature and a large number of species
mass fractions, particularly when detailed chemical mechanisms are considered. Directly reconstructing all of these scalar fields increases the
dimensionality of the super-resolution problem. Alternatively, selecting only a small
number of thermochemical scalars reduces the dimensionality, but may exclude information
contained in the remaining correlated quantities. In the present framework, principal
component analysis (PCA) is therefore used to obtain a compact and low-dimensional representation of the
thermochemical state. We have adopted a similar approach in our recent work implementing NIF for super-resolution, resulting in a computationally efficient reconstruction of species and temperatures~\citep{Amarathunga-ESSCI}. 

The retained principal components (PCs) combine information from the
original thermochemical scalars and allow the NFSR model to operate on a lower-dimensional
representation while retaining information associated with a larger set of scalar
quantities. The thermochemical scalar fields can subsequently be recovered from the reconstructed principal component fields through the inverse PCA transformation or, more accurately, through a nonlinear mapping using a simple artificial neural network (ANN), which has been shown to improve the reconstruction of
thermochemical scalars from a reduced set of PCs~\citep{Mirgolbabaei2013, PCA-linear}.

\begin{figure}[t]
        \centering
        \includegraphics[width=0.8\textwidth]{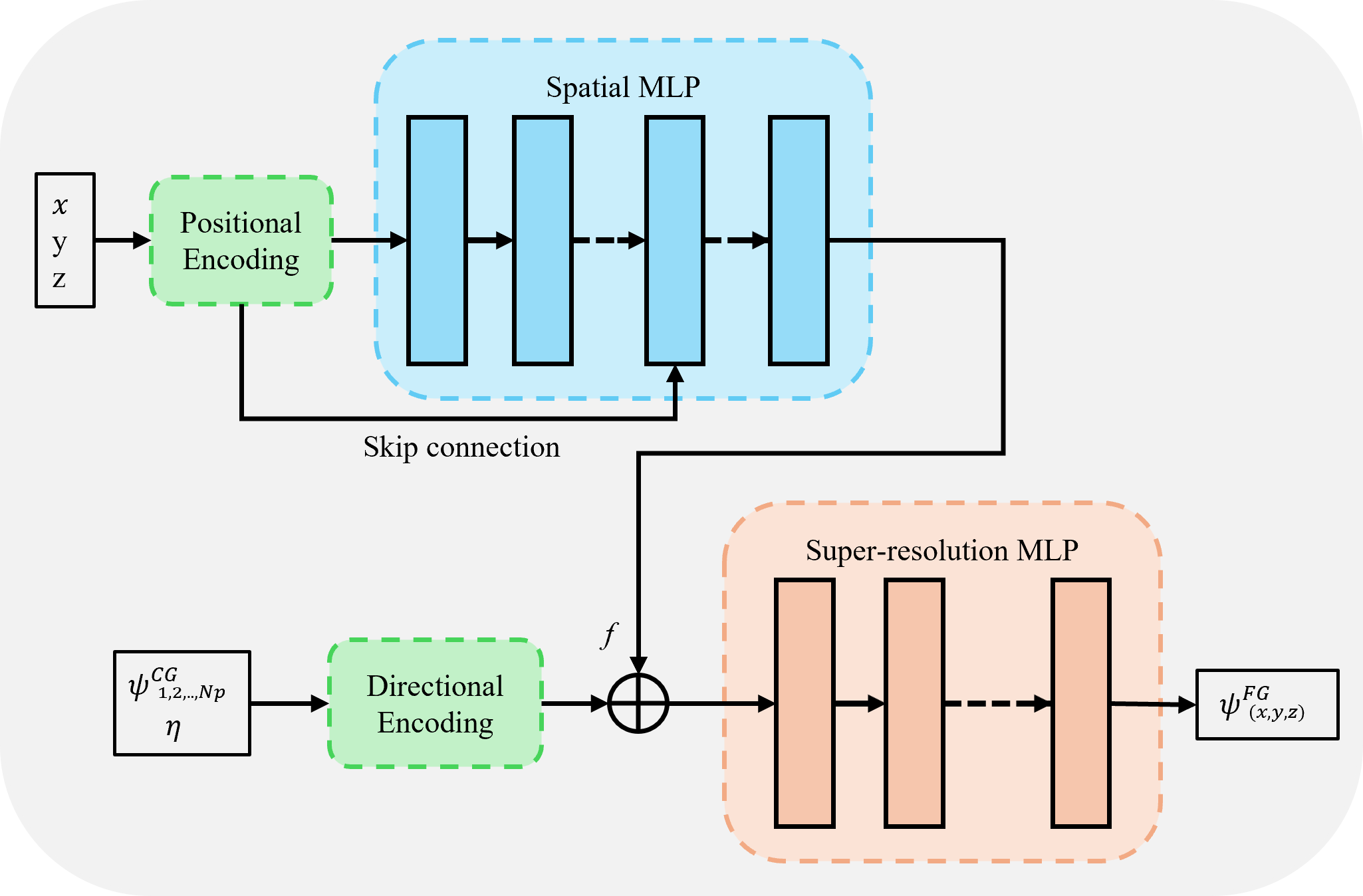}
        \caption{Proposed NeRF-inspired architecture for super-resolution (NFSR).}
        \label{fig:2}
\end{figure}

Converting the NeRF architecture to a super-resolution architecture also requires a modification to the input and output information of the network. The present framework adopts the coordinate-based representation, positional encoding, skip connection, and spatial-feature conditioning used in NeRF. However, important changes are introduced to adapt the architecture for the super-resolution of three-dimensional reacting-flow fields. Figure~\ref{fig:2} illustrates the proposed
NeRF-inspired architecture, hereafter referred to as the \emph{Neural Field
Super-Resolution} (NFSR) model. Unlike the original NeRF, the present model does not
predict volume density or radiance and does not implement ray projection or volume
rendering. Instead, the network directly reconstructs the fine-grid PC
fields at prescribed spatial coordinates using the corresponding local coarse-grid
PC information. The spatial coordinates, $\boldsymbol{x}=(x_1,x_2,x_3)=(x,y,z)$, are normalized within the local reconstruction region and transformed using positional encoding. For each
coordinate component $x_p$, where $p\in\{1,2,3\}$, the encoding is defined as
\begin{equation}
\gamma(x_p)
=
\left[
x_p,
\sin(2^{0}\pi x_p),
\cos(2^{0}\pi x_p),
\ldots,
\sin(2^{L-1}\pi x_p),
\cos(2^{L-1}\pi x_p)
\right],
\label{eq:positional_encoding}
\end{equation}
where $L$ denotes the number of frequency bands. The encoded spatial coordinates are
supplied to the spatial MLP, which generates the spatial feature vector,
$\boldsymbol{f}$. Similar to the original NeRF architecture, a skip connection is
implemented by concatenating the encoded coordinates with the output of an intermediate
layer of the spatial MLP. This provides the deeper layers of the network with direct
access to the spatial-coordinate information and facilitates the reconstruction of
localized features and strong spatial variations.

The viewing-direction input in the original NeRF architecture is replaced by a local
patch of coarse-grid PC values. For a target coordinate
$\boldsymbol{x}$, the local coarse-grid patch is represented by
\begin{equation}
\boldsymbol{P}_{\mathrm{CG}}(\boldsymbol{x})
=
\left[
\boldsymbol{\psi}^{\mathrm{CG}}_{1},
\boldsymbol{\psi}^{\mathrm{CG}}_{2},
\ldots,
\boldsymbol{\psi}^{\mathrm{CG}}_{N_p},
\boldsymbol{\eta}(\boldsymbol{x})
\right],
\label{eq:coarse_patch}
\end{equation}
where $\boldsymbol{\psi}^{\mathrm{CG}}_{i}$ denotes the vector of retained
PC values at the $i$th point of the local coarse-grid patch, and $N_p$
is the number of coarse-grid points included in the patch. The PC values
within this patch are flattened to form the conditioning input to the network. In addition,
the local fractional position of the target coordinate within its enclosing coarse-grid
cell is represented by
\begin{equation}
\boldsymbol{\eta}(\boldsymbol{x})
=
\left(
\eta_x,\eta_y,\eta_z
\right),
\qquad
\boldsymbol{\eta}\in[0,1]^3,
\label{eq:local_fraction}
\end{equation}
and appended to the flattened patch representation. The fractional-coordinate vector
enables the network to distinguish between fine-grid points that share the same
surrounding coarse-grid values by identifying the relative subcell position of each
target point.

As illustrated in Fig.~\ref{fig:2}, the coarse-grid conditioning vector is concatenated
with the spatial feature vector, $\boldsymbol{f}$, generated by the spatial MLP. The
combined representation is passed through the super-resolution head MLP, which predicts
the fine-grid PC solution at the queried coordinate. The proposed
mapping is therefore expressed as
\begin{equation}
\boldsymbol{\psi}^{\mathrm{FG}}(\boldsymbol{x})
=
G_{\boldsymbol{\Theta}}
\left(
\gamma(\boldsymbol{x}),
\boldsymbol{P}_{\mathrm{CG}}(\boldsymbol{x})
\right),
\label{eq:proposed_nerf_mapping}
\end{equation}
where $\boldsymbol{\psi}^{\mathrm{FG}}$ denotes the reconstructed fine-grid vector of
retained PCs. The corresponding fine-grid thermochemical scalar fields
are recovered from this representation using the inverse PCA transformation.

The main difference between the architectures in Figs.~\ref{fig:1} and~\ref{fig:2} is
therefore associated with the conditioning input and the predicted quantities. In the
original NeRF architecture, the directional input conditions the color-head MLP to
predict view-dependent radiance. In the present framework, the local coarse-grid
PC state and fractional position condition the super-resolution head
MLP to predict the corresponding fine-grid PC solution. Therefore, the
proposed model retains the continuous coordinate-conditioned representation of NeRF while
replacing radiance-field prediction and volume rendering with the direct pointwise
reconstruction of three-dimensional reacting-flow fields in the reduced PCA space.

\begin{figure}[t]
        \centering
        \includegraphics[width=0.8\textwidth]{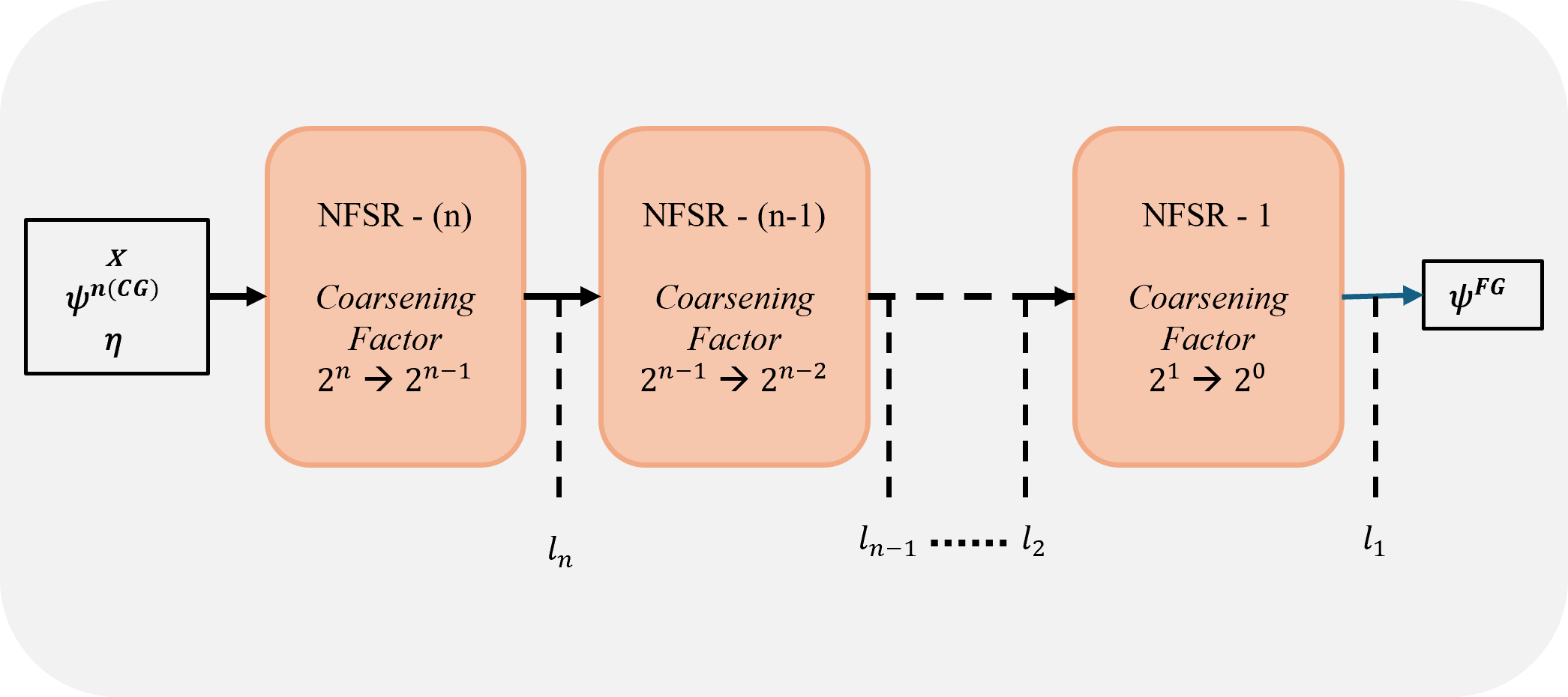}
        \caption{Hierarchical learning approach for super-resolution using the proposed model (NFSR).}
        \label{fig:3}
\end{figure}
\subsection{Hierarchical Learning Approach for Super-resolution}
\label{sec:hierarchical_framework}

Direct reconstruction of a fine-grid solution from a highly coarsened input requires the
model to recover a significant amount of spatial information in a single mapping. This
becomes increasingly challenging as the coarsening factor increases, particularly for
reacting-flow fields characterized by thin reaction zones and strong spatial variations.
To address this limitation, a hierarchical learning approach is implemented in
the proposed framework. Instead of directly reconstructing the fine-grid solution from the
coarsest input, the reconstruction is performed through a sequence of intermediate
resolution levels.

Figure~\ref{fig:3} illustrates the proposed hierarchical approach. The initial coarse-grid
field is defined using a coarsening factor of $2^n$ relative to the target fine-grid solution, although different factors may, in principle, be adopted.
The reconstruction then proceeds successively through coarsening factors
$2^{n-1},2^{n-2},\ldots,2^1$, until the fine-grid field corresponding to a coarsening factor
of $2^0=1$ is obtained. Therefore, each stage reduces the coarsening factor by a factor of
2 and increases the spatial resolution by a factor of 2 in each coordinate direction.

Let $\boldsymbol{\psi}^{(j)}$ denote the PC field corresponding to a
coarsening factor of $2^j$, where $j\in\{0,1,\ldots,n\}$. These resolution levels are
obtained from the reference fine-grid PC field through uniform spatial
subsampling, expressed as
\begin{equation}
\boldsymbol{\psi}^{(j)}
=
\mathcal{S}_{2^j}
\left(
\boldsymbol{\psi}^{\mathrm{FG}}
\right),
\qquad
j=0,1,\ldots,n,
\label{eq:hierarchical_levels}
\end{equation}
where $\mathcal{S}_{2^j}$ denotes uniform subsampling with a coarsening factor of $2^j$.
Accordingly,
\begin{equation}
\boldsymbol{\psi}^{(0)}
=
\boldsymbol{\psi}^{\mathrm{FG}},
\label{eq:fine_level}
\end{equation}
while $\boldsymbol{\psi}^{(n)}$ represents the coarsest input PC field.

The individual reconstruction models shown in Fig.~\ref{fig:3} retain the NFSR
architecture described in Section~\ref{sec:nerf_architecture}. The model at the $j$th
stage, denoted as NFSR-$j$, reconstructs a PC field with a coarsening
factor of $2^{j-1}$ from an input field with a coarsening factor of $2^j$. This stage-wise
mapping is expressed as
\begin{equation}
\widehat{\boldsymbol{\psi}}^{(j-1)}(\boldsymbol{x})
=
G_{\boldsymbol{\Theta}_j}
\left(
\gamma(\boldsymbol{x}),
\boldsymbol{P}^{(j)}(\boldsymbol{x})
\right),
\qquad
j=n,n-1,\ldots,1,
\label{eq:hierarchical_mapping}
\end{equation}
where $\boldsymbol{\Theta}_j$ denotes the trainable parameters of NFSR-$j$,
$\gamma(\boldsymbol{x})$ is the positional encoding of the queried spatial coordinate,
and $\boldsymbol{P}^{(j)}(\boldsymbol{x})$ is the local conditioning patch extracted
from the input PC field at level $j$.

For the first reconstruction stage, the conditioning patches are extracted
directly from the available coarse-grid PC field,
$\boldsymbol{\psi}^{(n)}$, corresponding to a coarsening factor of $2^n$.
For each subsequent stage, the conditioning patches are extracted from the
intermediate PC field predicted by the preceding NFSR model. The local
input at stage $j$ can therefore be represented by
\begin{equation}
\boldsymbol{P}^{(j)}(\boldsymbol{x})
=
\left[
\widehat{\boldsymbol{\psi}}^{(j)}_{1},
\widehat{\boldsymbol{\psi}}^{(j)}_{2},
\ldots,
\widehat{\boldsymbol{\psi}}^{(j)}_{N_p},
\boldsymbol{\eta}^{(j)}(\boldsymbol{x})
\right],
\label{eq:hierarchical_patch}
\end{equation}
where $\widehat{\boldsymbol{\psi}}^{(j)}_{i}$ denotes the predicted vector of retained
PCs at the $i$th point of the local patch, and
$\boldsymbol{\eta}^{(j)}(\boldsymbol{x})$ denotes the fractional position of the target
coordinate within the corresponding grid cell at level $j$. Consequently, the
intermediate prediction at each stage provides the coarse-grid information required by
the following stage.

As illustrated in Fig.~\ref{fig:3}, the prediction from each stage is compared with the
reference PC field at the corresponding resolution. For NFSR-$j$, the
reference target is the field $\boldsymbol{\psi}^{(j-1)}$, which has a coarsening factor
of $2^{j-1}$. The mean squared error associated with this stage is defined as
\begin{equation}
\ell_j
=
\frac{1}{N_jN_{\psi}}
\sum_{i=1}^{N_j}
\sum_{k=1}^{N_{\psi}}
\left(
\widehat{\psi}^{(j-1)}_{k,i}
-
\psi^{(j-1)}_{k,i}
\right)^2,
\qquad
j=n,n-1,\ldots,1,
\label{eq:stage_loss}
\end{equation}

where $N_j$ denotes the number of queried spatial points used at the output resolution of
the $j$th stage, $N_{\psi}$ is the number of retained PCs,
$\widehat{\psi}^{(j-1)}_{k,i}$ is the predicted value of the $k$th PC
at the $i$th spatial point, and $\psi^{(j-1)}_{k,i}$ is the corresponding reference value.

The complete hierarchical framework is trained by minimizing a weighted summation of
the losses from all reconstruction stages. The total loss is expressed as
\begin{equation}
\mathcal{L}_{\mathrm{total}}
=
\sum_{j=1}^{n}
\lambda_j\ell_j,
\label{eq:hierarchical_total_loss}
\end{equation}
where $\lambda_j$ denotes the prescribed weight of the $j$th stage to the total
loss. The weights satisfy
\begin{equation}
\lambda_j\geq 0,
\qquad
\sum_{j=1}^{n}\lambda_j=1.
\label{eq:loss_weight_constraint}
\end{equation}

The intermediate losses, $\ell_n,\ell_{n-1},\ldots,\ell_1$, provide direct supervision at
all resolution levels rather than evaluating the error only at the final fine-grid output.
Therefore, each NFSR model is constrained to accurately reconstruct the spatial structures
associated with its corresponding change in resolution. The final NFSR-1 model maps the
PC field with a coarsening factor of $2^1$ to the fine-grid solution with
a coarsening factor of $2^0$, yielding the final prediction,
$\widehat{\boldsymbol{\psi}}^{\mathrm{FG}}$. The corresponding fine-grid thermochemical
scalar fields can subsequently be recovered from the reconstructed PC
fields using the reconstruction procedure described in Section~\ref{sec:nerf_architecture}.

\subsection{Data Preparation and Model Training}

\subsubsection{Dataset description -- Premixed Flame H$_2$--Air DNS}
\label{sec:datadescription}

The proposed NFSR framework is applied to a
three-dimensional turbulent reacting-flow dataset from the BLASTNet data collection~\citep{blastnet3.0_2026}. The
selected configuration corresponds to a lean turbulent premixed hydrogen--air flame stabilized in
a slot burner geometry. The simulations are performed at atmospheric pressure with an unburned gas temperature of 300 K. A
premixed H$_2$--air mixture is injected through the central slot with a bulk velocity of
24 m/s and a turbulence intensity
$u'=2.4~\mathrm{m\,s^{-1}}$. The central jet is surrounded by two laminar coflows of
burned gas with a bulk velocity of 3.6 m/s.

Turbulence at the central inlet is generated using homogeneous and isotropic turbulence
based on a Passot--Pouquet spectrum~\citep{Passot_Pouquet_1987} with an integral length scale of
$l_t=2~\mathrm{mm}$. The BLASTNet database contains this configuration at several global
equivalence ratios. In the present work, the leanest configuration,
$\phi_g=0.35$, is selected to evaluate the reconstruction of highly localized
thermochemical structures associated with lean hydrogen combustion. For this case, the
DNS field is represented on a uniform Cartesian grid,
$951\times401\times201$, with a grid spacing of
$80~\mu\mathrm{m}$, and the streamwise domain length is
$L_x=76~\mathrm{mm}$.

The chemical kinetics of hydrogen oxidation are represented using the San Diego
chemical mechanism of ~\cite{saxena2006testing},
comprising nine species and 21 elementary reactions. The
available solution variables consist of the three velocity components, pressure,
temperature, density, and the species mass fractions of
H$_2$, H, O$_2$, OH, O, H$_2$O, HO$_2$, H$_2$O$_2$ and N$_2$.
For the present super-resolution problem, the DNS thermochemical fields provide the
reference fine-grid solution, while the corresponding coarse and intermediate resolution
fields are generated by the spatial subsampling procedure introduced in Section~2.2.

The $\phi_g=0.35$ dataset contains five statistically stationary snapshots separated by
$2~\mathrm{ms}$. In the present study, three snapshots are used for training, one snapshot
is used for validation, while the remaining snapshot is retained as an unseen test
realization. This separation ensures that the super-resolution performance is evaluated
on a turbulent flame realization that is not included during model training or model
selection. The resulting dataset is used to assess the ability of the proposed hierarchical
NFSR framework to reconstruct fine-scale reacting-flow structures from spatially
coarsened fields. However, owing to the limited number of available snapshots, the
reacting-flow dataset does not provide sufficient coverage of the statistical variability
even within the selected $\phi_g=0.35$ configuration. Consequently, the held-out snapshot
should be interpreted as an independent test realization rather than as a rigorously
defined test set representative of the full distribution of this operating condition.

Before data preparation, the computational domain is clipped to reduce the size of the
dataset while retaining the primary flame structure and the surrounding regions required
to characterize its spatial development. The domain is clipped symmetrically in the
$y$- and $z$-directions, while in the $x$-direction, which corresponds to the primary
flame-development direction, the far-field region downstream of the flame is truncated.
Following this operation, the original $951\times401\times201$ domain is reduced to
$641\times321\times193$ grid points, corresponding to a physical domain of approximately
$51.2\times25.6\times15.36~\mathrm{mm}^3$.

For data preparation, temperature and the eight species mass fractions, excluding N$_2$,
are selected. Each scalar is first normalized between 0 and 1 using its corresponding
minimum and maximum values. PCA is then applied to the normalized thermochemical state
to obtain a reduced representation that retains the dominant variations in the original
scalar fields. The PCs are selected such that at least $99\%$ of the
cumulative variance is retained. For the present dataset, this criterion results in 4
PCs, $(\boldsymbol{\psi}_1,\boldsymbol{\psi}_2,\boldsymbol{\psi}_3,\boldsymbol{\psi}_4)$, which together account for $99.57\%$ of the cumulative variance.

Once the PCs are obtained, an additional preprocessing step is applied
before model training. First, a constant shift is introduced to ensure that all PC values are positive. The shifted values are then raised to the one-fifth power,
which compresses the dynamic range and reduces the spread of the PC
data. The transformed PCs are subsequently standardized using their
respective mean and standard deviation before being used for the training and evaluation
of the NFSR framework.

Following the preprocessing of the PC fields, the computational domain is
decomposed into a set of local three-dimensional subdomains or blocks. This subdomain-based
learning strategy follows the approach adopted in the NIF-based
super-resolution framework~\citep{Amarathunga-ESSCI}, where the global domain was
partitioned into localized regions and each region was expressed in a normalized local coordinate system. This methodology enabled the model to learn spatial features based on their relative local structure rather than their absolute position in the computational domain. Such a representation promotes the learning of reusable local features that may occur at different spatial locations and under different flow conditions, thereby improving the potential for generalization. Although this domain decomposition methodology~\citep{Amarathunga-ESSCI} was demonstrated for
two-dimensional reacting-flow fields, the same localized learning principle is extended here to the three-dimensional reconstruction problem.

For the present three-dimensional implementation, each local block contains a coarse-grid representation of $5\times5\times5$ grid points, which
corresponds to $4\times4\times4$ coarse cells. With the overall coarsening factor of 8, the corresponding fine-grid subdomain contains $33\times33\times33$ grid points, or $32\times32\times32$ fine cells. For hierarchical reconstruction, the intermediate resolution levels therefore consist of the points of the grids $9\times9\times9$ and $17\times17\times17$ before reaching the final representation of the fine-grid $33\times33\times33$.

Within each block, the spatial coordinates are mapped to a local normalized coordinate system spanning $[-1,1]$ in each spatial direction. This local representation enables the NFSR model to learn the spatial structure within the subdomain independently of its absolute position in the global computational domain. A similar localized learning strategy was adopted in the NIF-based super-resolution framework for two-dimensional
reacting flows~\cite{Amarathunga-ESSCI}. In the present three-dimensional implementation, neighboring blocks are extracted with an
overlap ratio of $50\%$, such that adjacent subdomains share half of their spatial extent
in each overlapping direction. This overlap increases the diversity of the training data by
allowing similar local flow structures to appear at different relative positions within the blocks and by exposing the model to different combinations of neighboring local features.
As a result, the model is encouraged to learn the underlying local spatial patterns rather
than their specific locations within a given block.

\subsubsection{Data augmentation in training}

To further enrich the flame training data, a data augmentation strategy similar to the one adopted by ~\citet{chung2023} is used. In this approach, three-dimensional fields are subjected to discrete rotations and reflections.
The augmentation is applied stochastically
at the batch level during training. For each training batch, a rotation axis is selected
randomly from the $x$-, $y$-, and $z$-directions, while the rotation angle is selected
with equal probability from $0^{\circ}$, $90^{\circ}$, $180^{\circ}$, and
$270^{\circ}$. Thus, each rotation angle has a probability of $25\%$. Following the
rotation operation, a reflection is applied with a probability of $50\%$, with the
reflection axis randomly selected from the $x$-, $y$-, or $z$-direction. The selected
transformations are applied to the three-dimensional fields within the batch directly
through grid-index operations, avoiding spatial interpolation and the associated
modification of the PC fields.

The rotation and reflection operations expose the model to similar local flame structures appearing at different orientations within the computational domain. Although the present implementation differs slightly from that of ~\citet{chung2023}, the overall augmentation strategy follows the same objective of increasing the diversity of spatial structures available during training. This is particularly important for reacting flows, where flame fronts and localized high-gradient regions can occur over a wide range of orientations.

In addition to the spatial augmentation, a biased sampling strategy is introduced in the present framework to increase the frequency at which regions with strong spatial variations are selected during training. For each combination of snapshot and local coarse-grid window, the population standard deviation of the selected PC fields is evaluated and averaged to provide a measure of the local variation. The windows are then ranked according to this measure, such that regions with stronger variations are assigned larger sampling weights. The sampling weights are distributed between 1 and a prescribed bias ratio, $r_b$, according to
\begin{equation}
w_m
=
1+
(r_b-1)
\frac{R_m-1}{N_s-1},
\label{eq:bias_sampling_weight}
\end{equation}
where $R_m$ denotes the rank of the $m$th training sample and $N_s$ is the total number
of snapshot--window combinations. The corresponding sampling probability is obtained
by normalizing these weights over the complete training population. This biased sampling does not remove low-variation regions from the training data.
Instead, it increases the probability of selecting regions with stronger spatial variations,
which are more likely to contain flame fronts and localized reaction zone structures.

\subsubsection{Model training setup}

To evaluate the effects of the hierarchical reconstruction, biased sampling, and loss
function, four NFSR models are trained. The hierarchical NFSR (H-NFSR) model without
biased sampling is selected as the baseline configuration. The second model retains the
same H-NFSR architecture while introducing the biased sampling strategy described in
Section~2.3.2. The third model removes the intermediate reconstruction stages and directly
maps the coarsened input field to the fine-grid solution using a single-stage NFSR model.
Finally, the fourth model retains the baseline H-NFSR architecture but replaces the mean
squared error (MSE) loss with the mean absolute error (MAE). These configurations allow
the effects of the hierarchical reconstruction, sampling strategy, and loss criterion to be
evaluated independently.

The main model configurations are summarized in Table~\ref{tab:model_training}. The
H-NFSR models use three successive reconstruction stages corresponding to coarsening
factors of $8\rightarrow4\rightarrow2\rightarrow1$, with loss weights of
$(0.2,0.3,0.5)$ from the coarsest to the finest reconstruction stage. At each stage, the
local conditioning input is constructed from a $2\times2\times2$ grid-point patch
associated with the queried location. Each H-NFSR stage consists of four trunk layers and
six head layers with 256 neurons per hidden layer. The single-stage NFSR directly performs
the $8\rightarrow1$ reconstruction and uses a deeper network with 257 neurons per hidden
layer to maintain approximately the same total number of trainable parameters as the
H-NFSR models.

\begin{table}
\centering
\caption{Model configurations used for NFSR training and comparison.}
\label{tab:model_training}
\begin{tabular}{lcccc}
\hline
Configuration &
H-NFSR &
H-NFSR &
NFSR &
H-NFSR \\
&
no bias &
bias &
no bias &
MAE \\
&
(Base case) &
&
&
\\
\hline

Resolution levels &
$8,4,2,1$ &
$8,4,2,1$ &
$8,1$ &
$8,4,2,1$ \\

Bias sampling &
No &
Yes &
No &
No \\

Loss function &
MSE &
MSE &
MSE &
MAE \\

Loss weights &
$(0.2,0.3,0.5)$ &
$(0.2,0.3,0.5)$ &
$(1.0)$ &
$(0.2,0.3,0.5)$ \\

Hidden neurons &
256 &
256 &
257 &
256 \\

Trunk layers &
4 &
4 &
12 &
4 \\

Head layers &
6 &
6 &
20 &
6 \\

Skip layer &
2 &
2 &
6 &
2 \\

Total parameters &
2,190,348 &
2,190,348 &
2,194,270 &
2,190,348 \\

Best validation iteration &
330,000 &
194,000 &
112,000 &
564,000 \\

Time to best validation (h) &
6.97 &
5.83 &
0.50 &
8.62 \\

Total training time (600k iter.) (h) &
12.68 &
18.02 &
2.7 &
9.17 \\

\hline
\end{tabular}
\end{table}

All models use ReLU activation, eight positional-encoding frequencies for the spatial
coordinates, and two encoding frequencies for the conditioning input. The same training,
validation, and testing datasets and the same preprocessing procedure are used for all
configurations. A coarse-grid subdomain size of $5\times5\times5$ points and an overlap
ratio of $50\%$ are maintained throughout the model comparisons.

Training is performed for $600,000$ iterations using an exponentially decaying learning
rate. The initial learning rate is set to $10^{-3}$ and is reduced by multiplying this rate by a factor of
0.9 every $15,000$ iterations, with a minimum learning rate of $10^{-6}$. The
training and evaluation batches contain 32,784 and 8,196 queried spatial points,
respectively. Training is performed using four NVIDIA RTX A6000 GPUs, each with
48~GB of memory. The reported training times in Table~\ref{tab:model_training} are based on this hardwaer. For each configuration, the model parameters corresponding to the
minimum validation loss are retained for the subsequent reconstruction and comparison.

\subsection{Evaluation metrics}
\label{sec:evaluation_metrics}

The reconstruction performance is evaluated using three complementary metrics: the
normalized mean absolute error (NMAE), the normalized maximum absolute error (MaxAbs),
and the Structural Similarity Index Measure (SSIM). The NMAE and MaxAbs quantify the
pointwise reconstruction accuracy, while SSIM provides an additional assessment of the
preservation of spatial structures within the reconstructed fields.

For the $k$th PC, let $\psi_{k,i}$ and
$\widehat{\psi}_{k,i}$ denote the reference and reconstructed values, respectively, at
the $i$th spatial location. The global range of the corresponding PC is
defined as
\begin{equation}
R_k
=
\psi_{k,\max}-\psi_{k,\min},
\label{eq:global_range}
\end{equation}
where $\psi_{k,\max}$ and $\psi_{k,\min}$ denote the maximum and minimum values of the
reference PC field. The percentage NMAE is then defined as
\begin{equation}
\mathrm{NMAE}_k
=
\frac{100}{R_k}
\left[
\frac{1}{N}
\sum_{i=1}^{N}
\left|
\widehat{\psi}_{k,i}-\psi_{k,i}
\right|
\right],
\label{eq:nmae}
\end{equation}
where $N$ is the total number of evaluated spatial points. The percentage maximum
absolute error is defined as
\begin{equation}
\mathrm{MaxAbs}_k
=
\frac{100}{R_k}
\max_i
\left|
\widehat{\psi}_{k,i}-\psi_{k,i}
\right|.
\label{eq:maxabs}
\end{equation}

To evaluate the structural fidelity of the reconstructed fields, SSIM is also employed
\citep{wang2004ssim}. Unlike pointwise error measures, SSIM compares the local mean,
variance, and covariance of the reconstructed and reference fields and therefore provides
a measure of how well the spatial organization and local structures are preserved. This
property is particularly useful for super-resolution of reacting flows, where reconstruction
quality depends not only on reducing pointwise errors but also on recovering thin
reaction-zone structures, strong gradients, and localized thermochemical features. SSIM
has also been adopted for the evaluation of turbulent-flow super-resolution because of its
ability to assess both structural and statistical similarities between reconstructed and
reference fields~\citep{chung2023}.

The SSIM between two local regions, $x$ and $y$, is defined as
\begin{equation}
\mathrm{SSIM}(x,y)
=
\frac{
(2\mu_x\mu_y+C_1)
(2\sigma_{xy}+C_2)
}{
(\mu_x^2+\mu_y^2+C_1)
(\sigma_x^2+\sigma_y^2+C_2)
},
\label{eq:ssim}
\end{equation}
where $\mu_x$ and $\mu_y$ denote the local means,
$\sigma_x^2$ and $\sigma_y^2$ are the corresponding variances, and
$\sigma_{xy}$ is their covariance. The stabilization constants are defined as
\begin{equation}
C_1=(K_1R_k)^2,
\qquad
C_2=(K_2R_k)^2,
\label{eq:ssim_constants}
\end{equation}
where $K_1=0.01$ and $K_2=0.03$. In the present study, SSIM is evaluated using a
$9\times9\times9$ local window over the three-dimensional PC fields.

For each dataset split, NMAE and MaxAbs are evaluated over all spatial points and
snapshots contained within the corresponding split. SSIM is evaluated independently for
each snapshot and PC, and the mean SSIM over the corresponding
snapshots is reported.

\section{Results and Discussion}
\label{sec:Results}

\subsection{Local and global reconstruction of premixed $H_2$-air DNS}
\label{sec:premixed_results}

\begin{figure}[t]
    \centering

    \includegraphics[width=\textwidth]{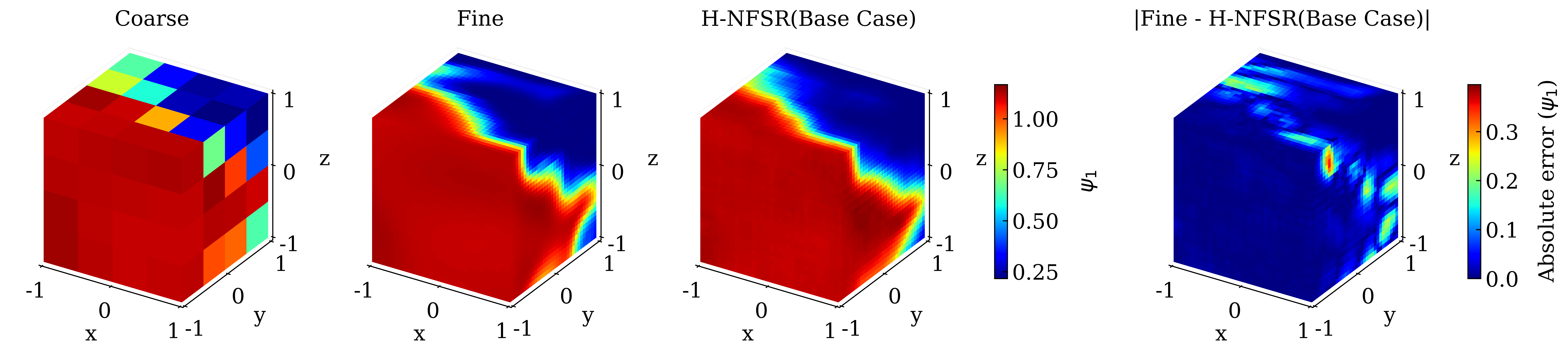}\\[2mm]
    \includegraphics[width=\textwidth]{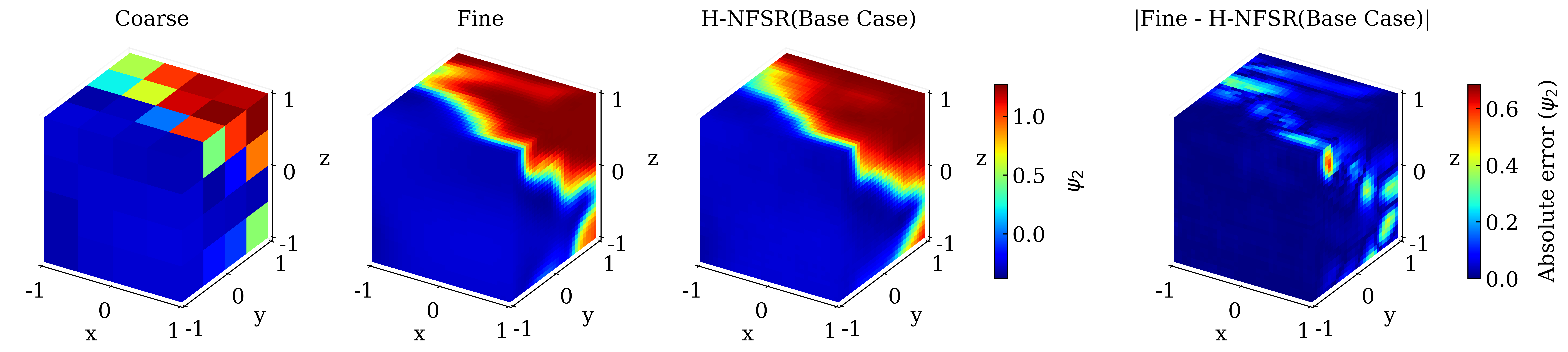}\\[2mm]
    \includegraphics[width=\textwidth]{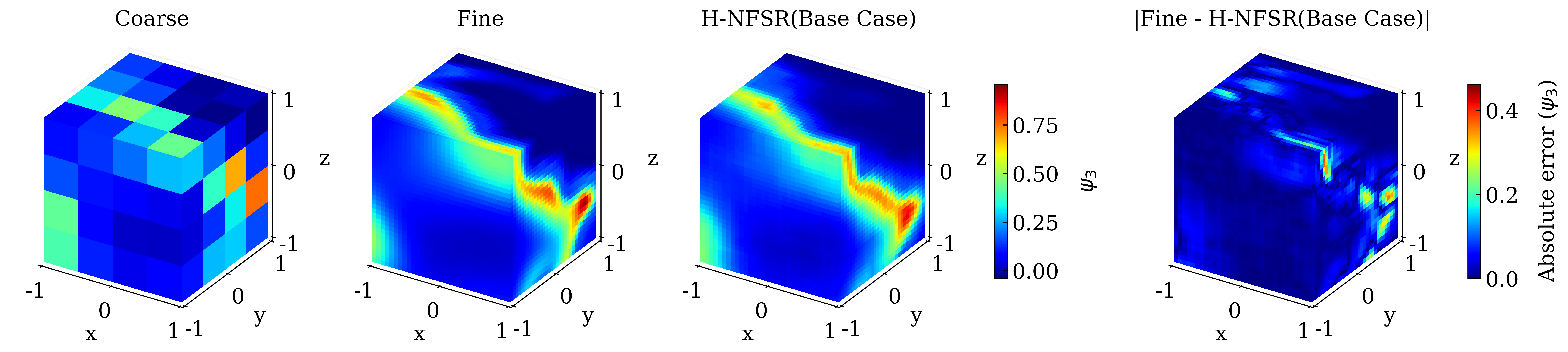}\\[2mm]
    \includegraphics[width=\textwidth]{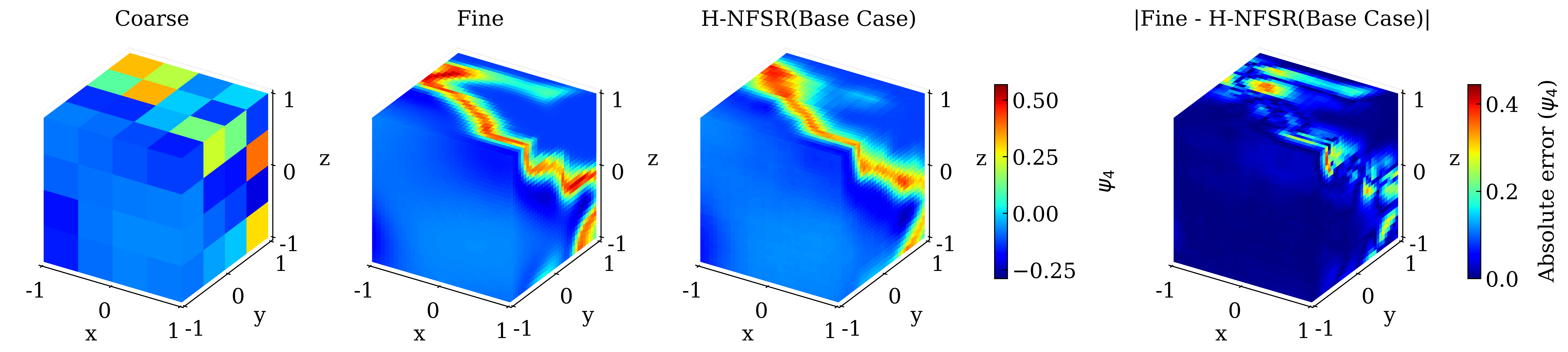}

    \caption{Representative local subdomain reconstruction using the baseline H-NFSR
    model for the 4 retained PCs. From top to bottom, the rows
    correspond to $\psi_1$, $\psi_2$, $\psi_3$, and $\psi_4$, respectively. For each
    PC, the coarse-grid input, reference fine-grid field, H-NFSR
    reconstruction, and absolute reconstruction error are shown. Coordinates are normalized between -1 and 1.}
    \label{fig:local_pc_reconstruction}
\end{figure}

The reconstruction performance of the proposed framework is first examined at the local
subdomain level. Figure~\ref{fig:local_pc_reconstruction} presents a representative
subdomain from the test dataset reconstructed using the baseline H-NFSR model. For each
of the 4 retained PCs, the corresponding coarse-grid input, reference
fine-grid field, H-NFSR reconstruction, and absolute reconstruction error are shown. The
coarse representation contains only $5\times5\times5$ grid points, while the target
fine-grid field contains $33\times33\times33$ grid points, corresponding to a coarsening
factor of eight in each spatial direction.

\begin{table}
\centering
\caption{Comparison of the baseline H-NFSR model and tricubic interpolation for the
validation and test datasets. NMAE and MaxAbs are reported as percentages.}
\label{tab:base_vs_tricubic}
\begin{tabular}{llccc|ccc}
\hline
& &
\multicolumn{3}{c|}{H-NFSR (base case)} &
\multicolumn{3}{c}{Tricubic interpolation} \\
Dataset & PC &
NMAE (\%) & MaxAbs (\%) & SSIM &
NMAE (\%) & MaxAbs (\%) & SSIM \\
\hline

Validation
& $\psi_1$ & 1.50 & 63.64 & 0.9530 & 1.75 & 58.53 & 0.9314 \\
& $\psi_2$ & 1.35 & 66.58 & 0.9340 & 1.71 & 62.86 & 0.9049 \\
& $\psi_3$ & 0.93 & 58.56 & 0.9393 & 1.35 & 68.86 & 0.8947 \\
& $\psi_4$ & 1.63 & 59.98 & 0.8730 & 2.39 & 61.35 & 0.7997 \\
\hline

Test
& $\psi_1$ & 1.50 & 72.09 & 0.9521 & 1.76 & 55.67 & 0.9297 \\
& $\psi_2$ & 1.36 & 73.68 & 0.9324 & 1.71 & 57.28 & 0.9033 \\
& $\psi_3$ & 0.95 & 65.81 & 0.9383 & 1.36 & 67.15 & 0.8928 \\
& $\psi_4$ & 1.63 & 57.86 & 0.8725 & 2.37 & 67.12 & 0.7997 \\
\hline
\end{tabular}
\end{table}

The H-NFSR predictions recover the major spatial structures that are not resolved in the
coarse input. For $\psi_1$ and $\psi_2$, the reconstructed fields closely follow the
large-scale transitions observed in the reference fine-grid solution while recovering the
sharper spatial variations across these regions. The reconstruction errors are mainly
concentrated near the strongest gradients, while the smoother regions exhibit considerably
smaller errors.

The reconstruction of $\psi_3$ and $\psi_4$ provides a more challenging assessment of
the framework. In particular, $\psi_4$ contains thin and highly localized structures that
are only sparsely represented in the coarse-grid input. Despite the substantial loss of
spatial information introduced by the coarsening, the H-NFSR model recovers the location
and overall shape of these structures with good agreement with the fine-grid reference.
Similar behavior is observed for $\psi_3$, where localized spatial variations that are
poorly represented by the coarse field are recovered in the reconstructed solution.
Although the largest errors remain concentrated around the strongest gradients, the model
recovers considerably more of the fine-scale structure than is directly available from the
coarse representation.

\begin{figure}[htbp]
    \centering
    \includegraphics[width=0.75\textwidth]{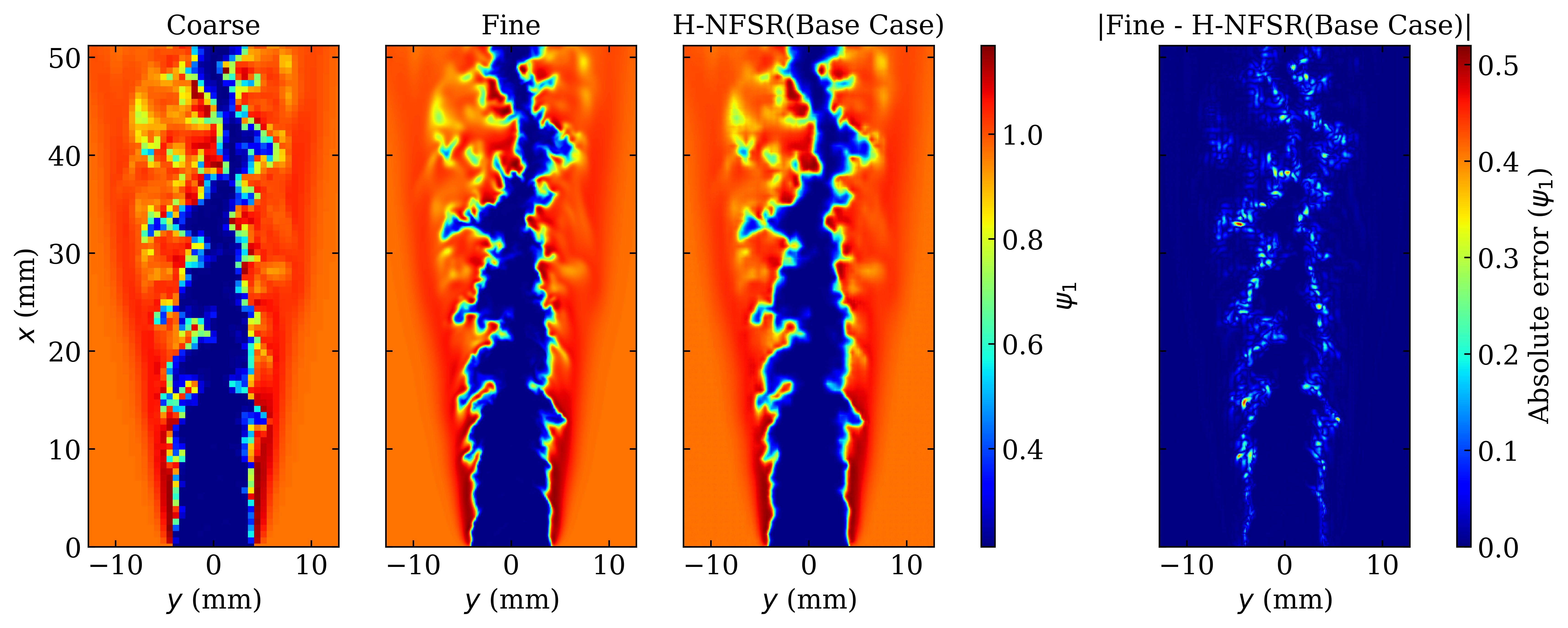}\\[1mm]
    \includegraphics[width=0.75\textwidth]{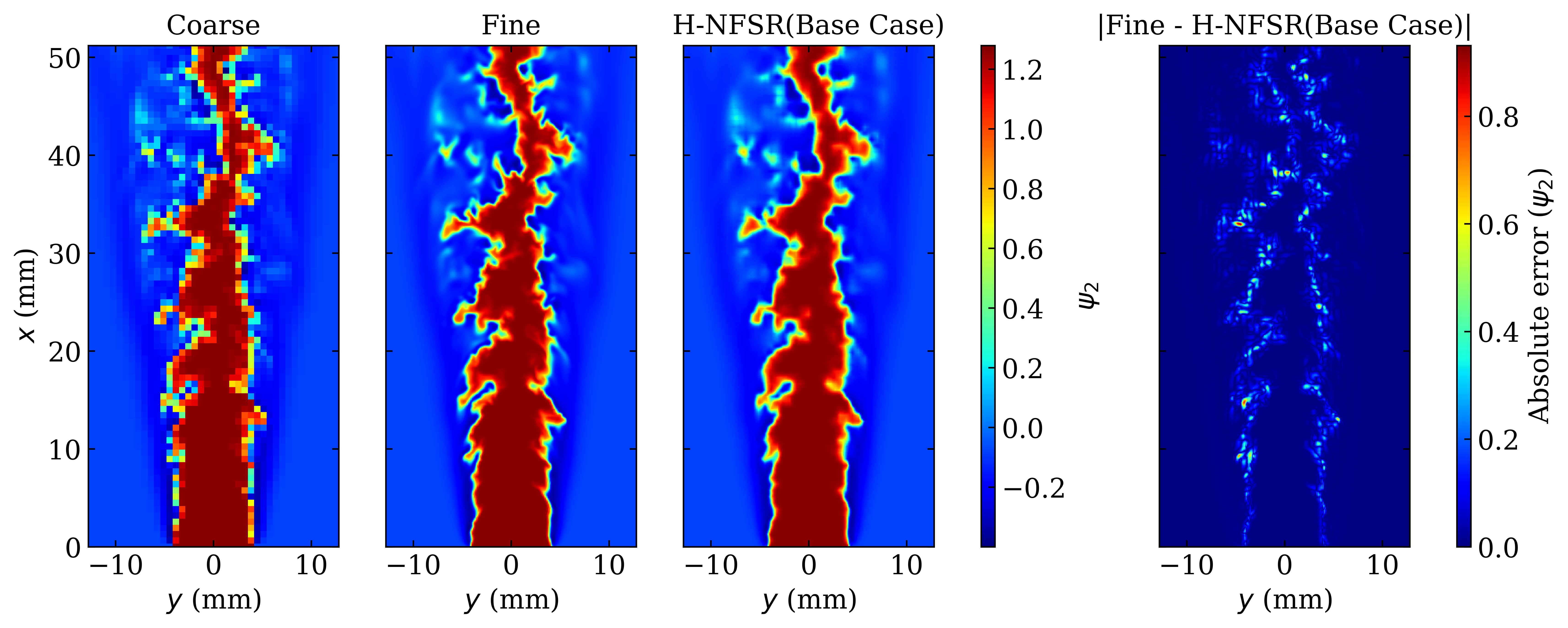}\\[1mm]
    \includegraphics[width=0.75\textwidth]{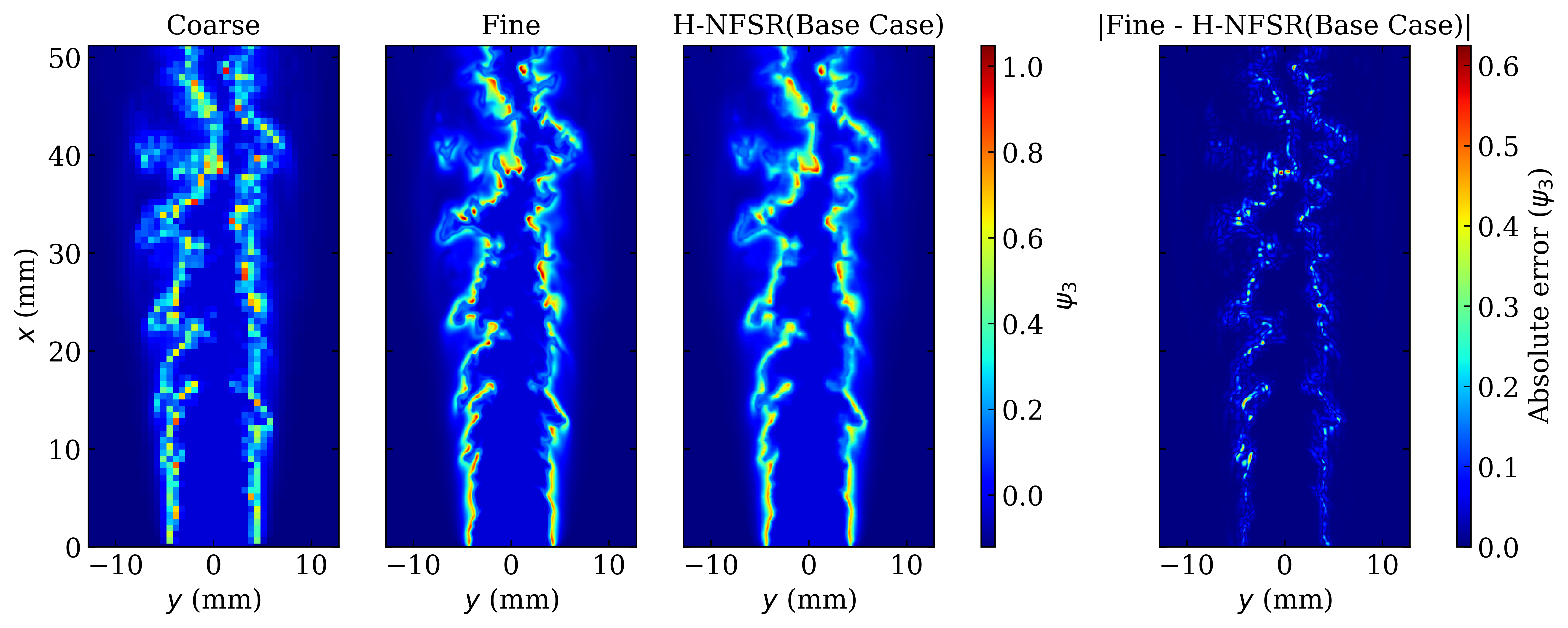}\\[1mm]
    \includegraphics[width=0.75\textwidth]{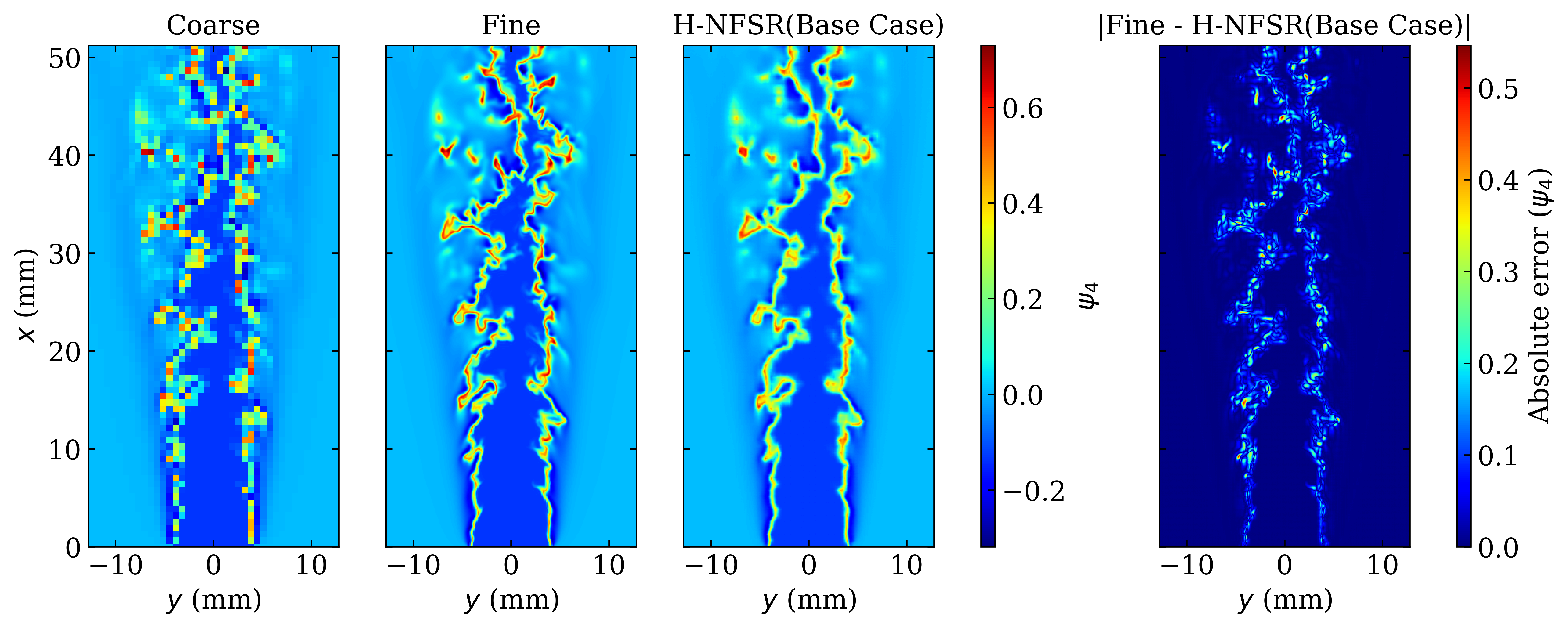}

    \caption{Representative two-dimensional $x-y$ slices at $z=0$ from the unseen test
    dataset using the baseline H-NFSR model. From top to bottom, the rows correspond to
    $\psi_1$, $\psi_2$, $\psi_3$, and $\psi_4$, respectively. For each PC,
    the coarse-grid field, reference fine-grid DNS field, H-NFSR reconstruction, and absolute
    reconstruction error are shown.}
    \label{fig:global_pc_reconstruction}
\end{figure}

\begin{figure}[t]
    \centering
    \includegraphics[width=0.92\textwidth]{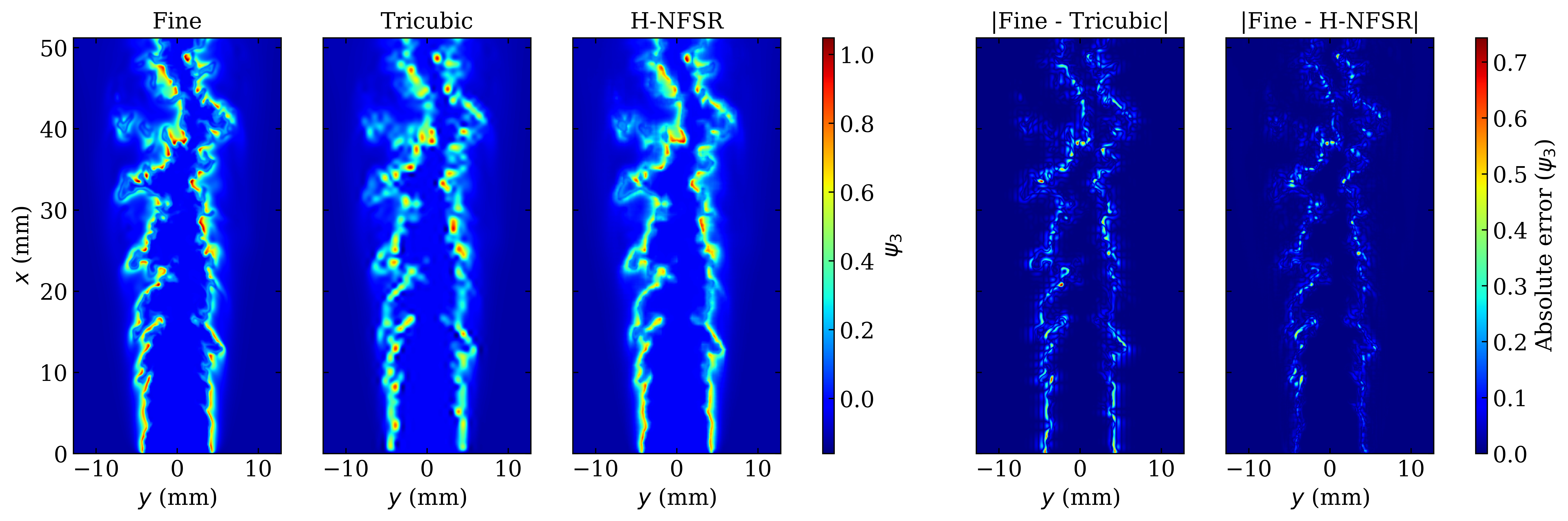}\\[2mm]
    \includegraphics[width=0.92\textwidth]{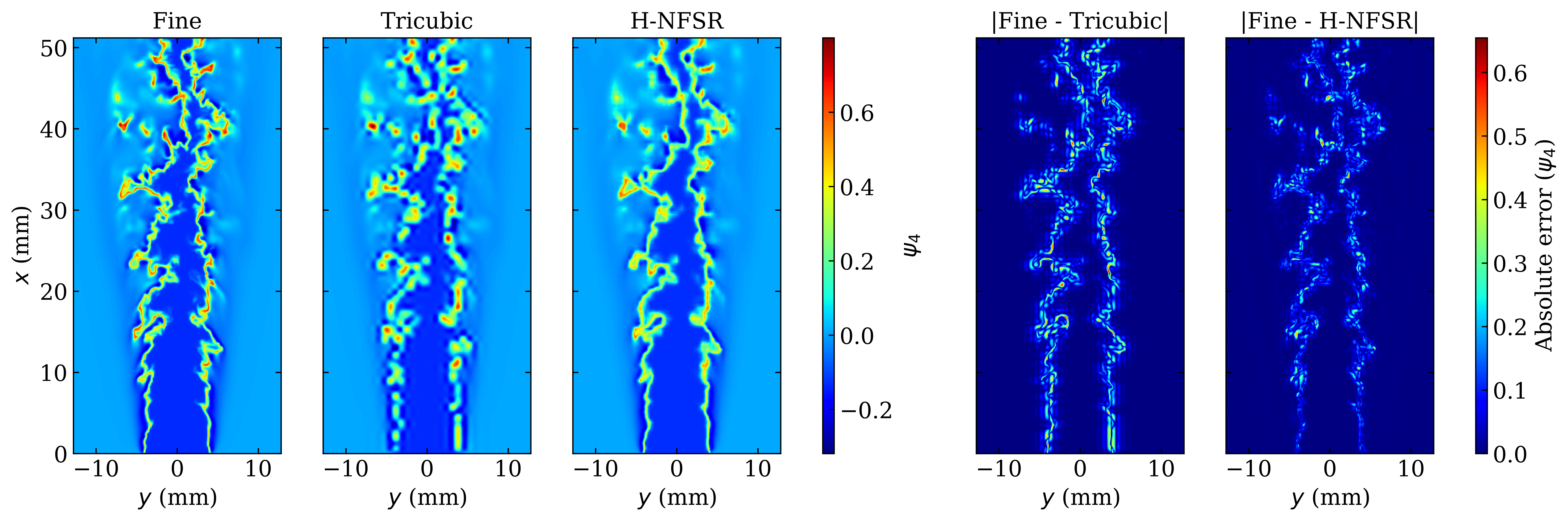}
    \caption{Comparison of tricubic interpolation and the baseline H-NFSR reconstruction
    for representative global $x-y$ slices at $z=0$ from the unseen test dataset.
    From top to bottom, the rows correspond to $\psi_4$ and $\psi_3$, respectively.
    For each PC, the reference fine-grid DNS field, tricubic
    interpolation, H-NFSR reconstruction, absolute error of the tricubic interpolation,
    and absolute error of the H-NFSR reconstruction are shown.}
    \label{fig:global_tricubic_comparison}
\end{figure}

To provide a non-learned reference for the reconstruction, the baseline H-NFSR model is
compared with tricubic interpolation using the same coarse-grid fields.
Table~\ref{tab:base_vs_tricubic} summarizes the reconstruction errors for the 4
PCs on the validation and test datasets. The H-NFSR model provides
lower NMAE and higher SSIM than tricubic interpolation for all 4 PCs
in both datasets. The improvement becomes particularly important for $\psi_3$ and
$\psi_4$, where a larger fraction of the field variation is associated with localized
spatial structures.

For the test dataset, the NMAE of $\psi_3$ decreases from $1.36\%$ with tricubic
interpolation to $0.95\%$ with H-NFSR, corresponding to a reduction of approximately
$30.4\%$. The corresponding SSIM increases from 0.8928 to 0.9383. A similar improvement
is observed for $\psi_4$, where the NMAE decreases from $2.37\%$ to $1.63\%$,
representing a reduction of approximately $31.1\%$, while the SSIM increases from
0.7997 to 0.8725. The improvement in $\psi_4$ is particularly important since the
representative field contains thin and highly localized structures that are only weakly
represented in the coarse-grid input. The H-NFSR reconstruction therefore recovers
additional spatial information beyond that obtained through direct tricubic interpolation.
Similar trends are observed for the validation dataset. 

The maximum absolute error shows a less uniform trend. For $\psi_3$ and $\psi_4$, the
H-NFSR model provides comparable or lower maximum errors than tricubic interpolation,
while larger isolated maximum errors are observed for $\psi_1$ and $\psi_2$ in the test
dataset. However, the consistently lower NMAE and higher SSIM indicate that the H-NFSR
model provides improved reconstruction over the majority of the field and more accurately
preserves the spatial structure of the PC fields. Therefore, the main
advantage of the learned reconstruction is the recovery of fine-scale spatial information
rather than simply a reduction of the largest isolated pointwise error.

Following the local reconstruction, the predictions from the overlapping subdomains are
assembled to recover the global PC fields over the clipped computational
domain. Since adjacent subdomains overlap by $50\%$, multiple local predictions are
available within shared regions. The overlapping subdomain predictions are assembled using the selection-based
procedure described by~\citet{Amarathunga-ESSCI}. For each subdomain, a coarse-grid
mean absolute error (MAE) is evaluated by comparing the reconstructed and reference
values at the corresponding coarse-grid locations. Within the overlapping regions, the
prediction associated with the subdomain having the lowest coarse-grid MAE is retained.
This procedure favors the locally more accurate prediction and is used to construct the
continuous fine-grid field over the full computational domain.

Figure~\ref{fig:global_pc_reconstruction} presents representative two-dimensional
$x-y$ slices at $z=0$ from the unseen test dataset for the 4 retained principal
components. For each PC, the coarse-grid field, reference fine-grid DNS
field, reconstructed H-NFSR field, and the corresponding absolute error are shown. The comparison indicates that the localized reconstruction capability observed at the subdomain
level is retained after assembling the predictions over the full computational domain and
when applied to a snapshot not used during model training or validation.

Across all 4 PCs, the H-NFSR reconstruction recovers substantially
more spatial detail than is available in the coarse-grid input. The coarse fields retain
only the broad large-scale distribution of the PC structures and do not
resolve the thin interfaces, localized pockets, and small-scale variations present in the
fine-grid DNS field. In contrast, the reconstructed H-NFSR fields recover the overall
shape of the flame-associated structures together with a significant portion of the
fine-scale wrinkling and localized variations.

This improvement is particularly evident for $\psi_3$ and $\psi_4$. These PCs contain thinner and more localized structures, which are only weakly
represented in the coarse-grid input. Despite this substantial loss of spatial information,
the H-NFSR model reconstructs the PCs' contours with good fidelity. The absolute errors remain primarily concentrated along
the sharpest gradients and thin interface regions, while the majority of the remaining
domain exhibits relatively small errors. The global reconstruction confirms that the H-NFSR framework can recover fine-scale spatial information beyond that
available directly from the coarse-grid representation for the unseen test realization.

To further demonstrate the reconstruction capability of the proposed framework, 
Fig.~\ref{fig:global_tricubic_comparison} compares the global $x-y$ slices at $z=0$
from the unseen test dataset for $\psi_3$ and $\psi_4$ using tricubic interpolation and
the baseline H-NFSR model. These 2 PCs are selected because they
contain thinner and more localized structures than $\psi_1$ and $\psi_2$, and therefore
provide a more stringent assessment of the reconstruction method.

For both $\psi_3$ and $\psi_4$, tricubic interpolation recovers only a smoothed
approximation of the fine-grid field and fails to reproduce a substantial portion of the
small-scale wrinkling and localized variations present in the DNS reference. In contrast,
the H-NFSR reconstruction recovers noticeably sharper interfaces and substantially more of
the fine-scale topology. This improvement is particularly evident for $\psi_4$, where the
reference field contains thin and fragmented structures that are only weakly represented
in the coarse-grid input. The H-NFSR reconstruction retains much more of these structures,
whereas tricubic interpolation smooths them out and loses a significant amount of the
spatial detail.

This qualitative improvement is in agreement with the quantitative results reported in
Table~\ref{tab:base_vs_tricubic}. For the test dataset, the NMAE of $\psi_3$ is reduced
from $1.36\%$ for tricubic interpolation to $0.95\%$ for H-NFSR, while the SSIM
increases from 0.8928 to 0.9383. For $\psi_4$, the NMAE is reduced from $2.37\%$ to
$1.63\%$, and the SSIM increases from 0.7997 to 0.8725. The agreement between the qualitative and
quantitative comparisons indicates that the proposed H-NFSR framework recovers
significantly more spatial detail than tricubic interpolation for the higher-order
PCs.

\subsection{Effect of Hierarchical Learning Approach}
\label{sec:hierarchical_effect}

\begin{figure}[htbp]
    \centering
    \includegraphics[width=\textwidth]{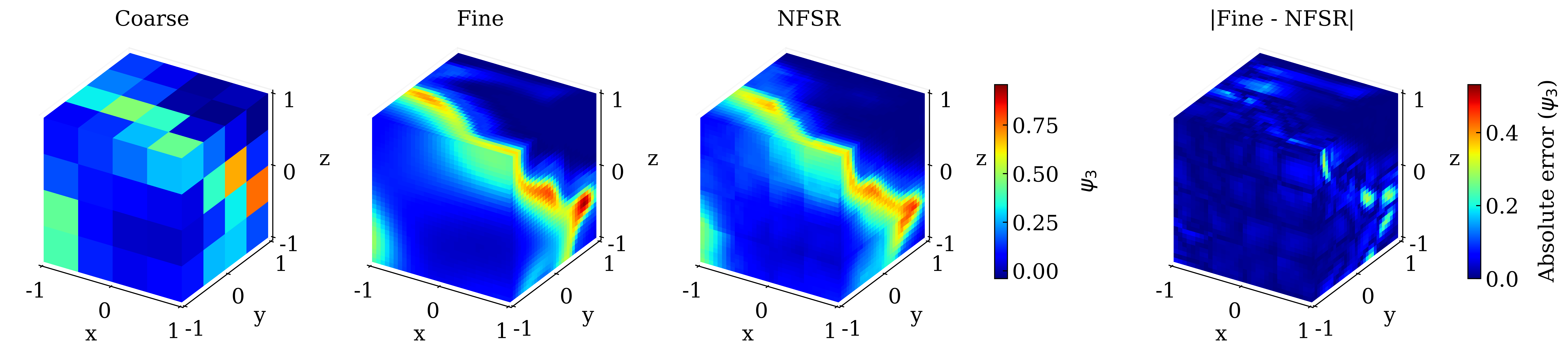}\\[2mm]
    \includegraphics[width=\textwidth]{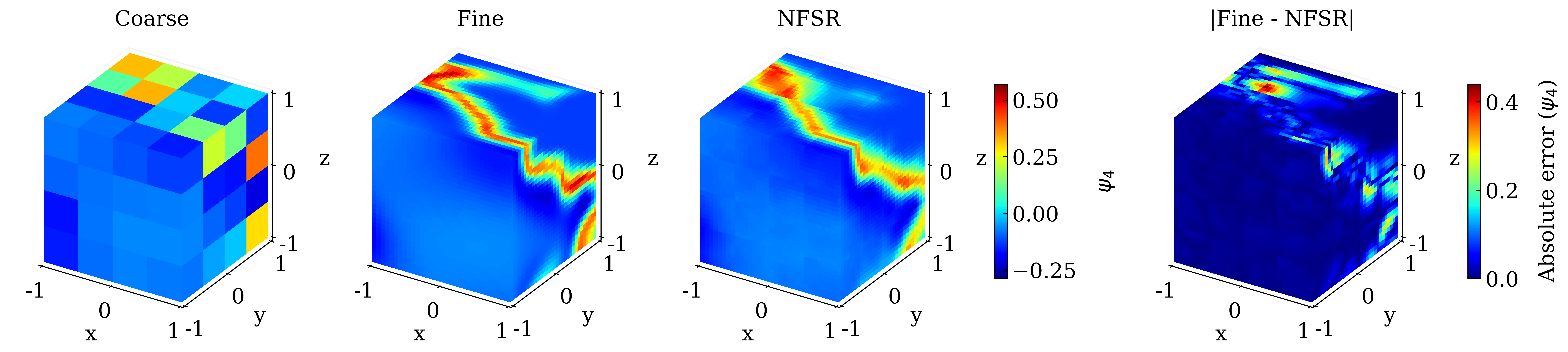}

    \caption{Representative local subdomain reconstructions from the unseen test
    dataset using the single-stage NFSR model for $\psi_3$ and $\psi_4$.
    From top to bottom, the rows correspond to $\psi_3$ and $\psi_4$, respectively.
    The reconstructed fields retain visible spatial imprinting associated with the
    underlying coarse-grid cells.}
    \label{fig:nonhier_local}
\end{figure}

\begin{figure}[htbp]
    \centering
    \includegraphics[width=0.92\textwidth]{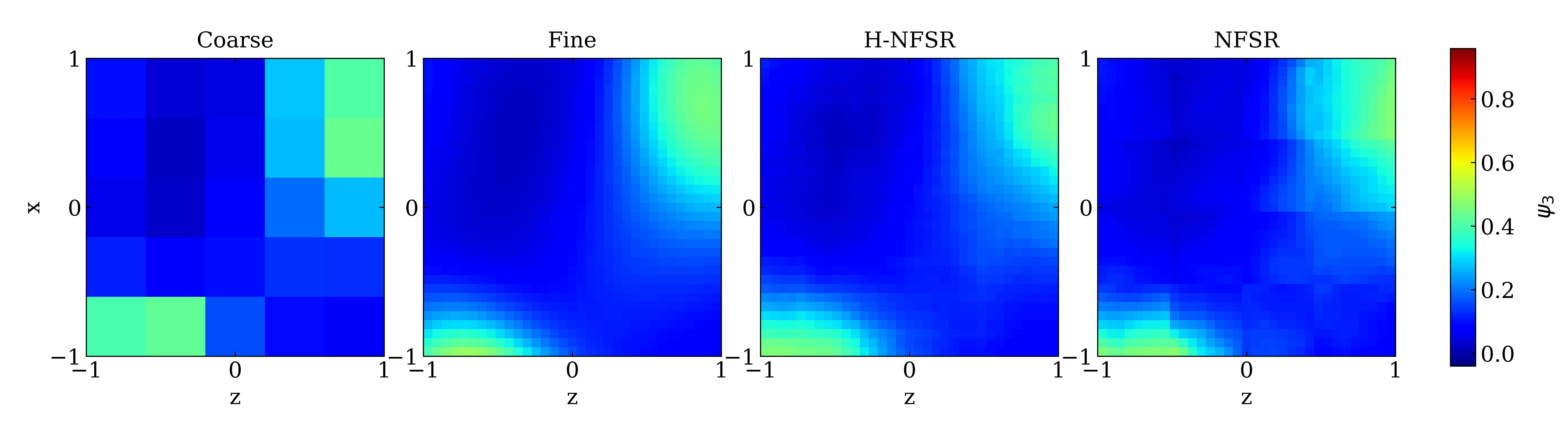}\\[2mm]
    \includegraphics[width=0.92\textwidth]{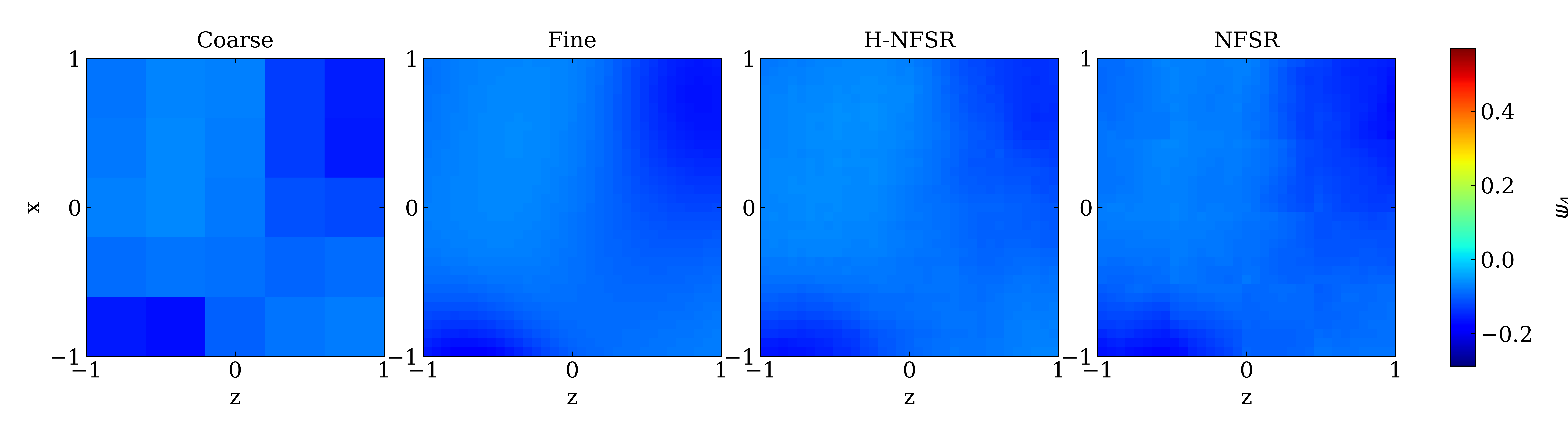}

    \caption{Comparison of the H-NFSR and single-stage NFSR reconstructions on a
    representative local $x$--$z$ face at $y=-1$ from the unseen test dataset.
    From top to bottom, the rows correspond to $\psi_3$ and $\psi_4$, respectively.
    For each PC, the coarse-grid field, reference fine-grid field, H-NFSR
    reconstruction, and single-stage NFSR reconstruction are shown. The H-NFSR
    reconstruction more closely follows the smooth spatial variation of the fine-grid
    reference, whereas the single-stage NFSR retains more visible imprinting of the
    underlying coarse-grid cells.}
    \label{fig:hier_nonhier_local_face}
\end{figure}

\begin{figure}[htbp]
    \centering
    \includegraphics[width=0.92\textwidth]{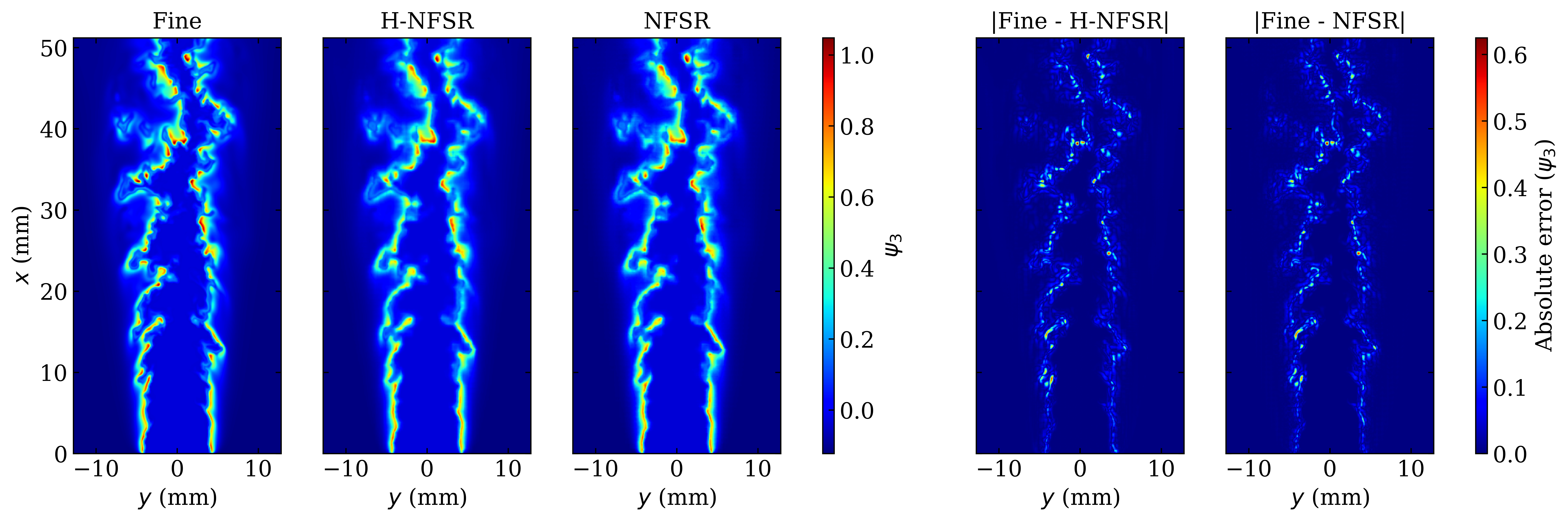}\\[2mm]
    \includegraphics[width=0.92\textwidth]{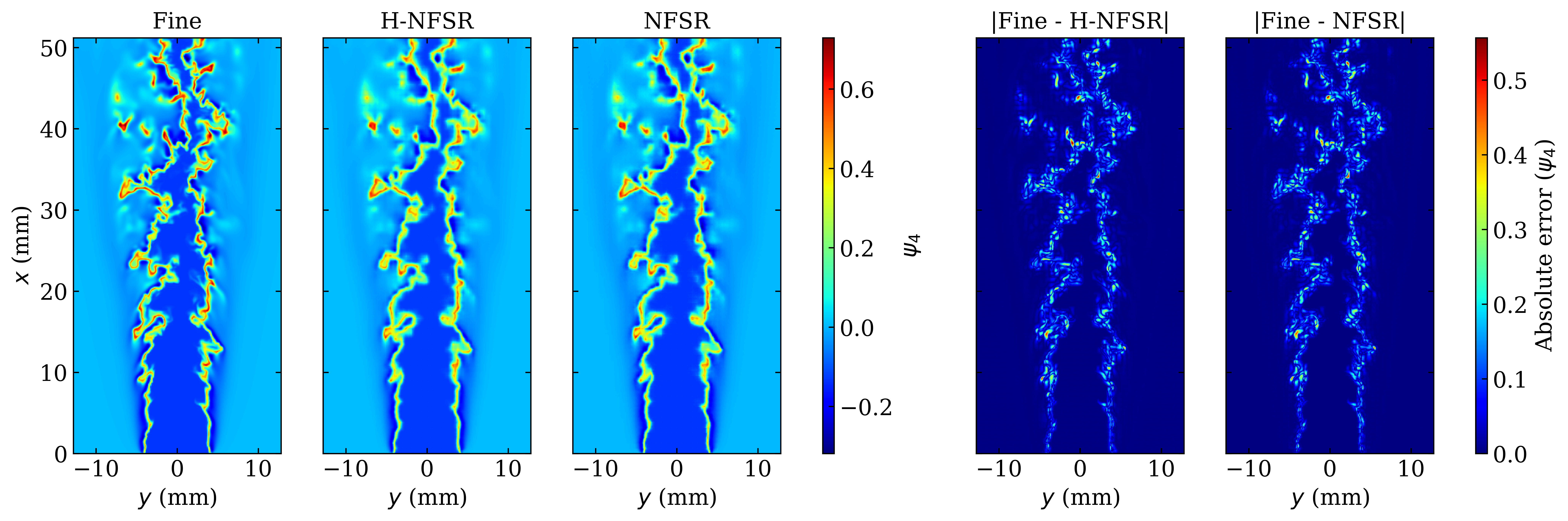}
    \caption{Comparison of the baseline H-NFSR and single-stage NFSR models for
    representative global $x-y$ slices at $z=0$ from the unseen test dataset.
    From top to bottom, the rows correspond to $\psi_3$ and $\psi_4$, respectively.
    For each PC, the reference fine-grid DNS field, H-NFSR
    reconstruction, NFSR reconstruction, and their corresponding absolute errors are
    shown.}
    \label{fig:hier_nonhier_global}
\end{figure}

The effect of the hierarchical reconstruction is evaluated by comparing the baseline
H-NFSR model with the single-stage NFSR model. Both models use approximately the same
number of trainable parameters and are trained without biased sampling using the same
training procedure. The main difference lies in the reconstruction strategy:
H-NFSR progressively reconstructs the field through $8\rightarrow4\rightarrow2\rightarrow1$,
whereas NFSR directly performs the $8\rightarrow1$ mapping.

Figure~\ref{fig:nonhier_local} shows representative local reconstructions from the
unseen test dataset for $\psi_3$ and $\psi_4$ using the single-stage NFSR model on a single block of the data. Although the direct NFSR model successfully recovers the main fine-grid structures,
visible imprinting of the original coarse-grid cells remains in the reconstructed fields.
These artifacts appear as weak block-like variations within the fine-grid prediction,
indicating that some dependence on the underlying coarse-grid representation is retained
after the direct $8\rightarrow1$ reconstruction.

The effect of the hierarchical reconstruction on these local artifacts is examined more
directly in Fig.~\ref{fig:hier_nonhier_local_face}, which compares the coarse-grid field,
fine-grid reference, H-NFSR reconstruction, and single-stage NFSR reconstruction on a
representative local $x$--$z$ face at $y=-1$ using a single block of the data . The difference is particularly noticeable
for $\psi_3$ and $\psi_4$, where the reference fields contain localized spatial variations.
Although both models recover the principal fine-grid structures, the single-stage NFSR
retains more visible imprinting of the underlying coarse-grid cells. In comparison, the
H-NFSR reconstruction exhibits smoother spatial variation and more closely follows the
fine-grid reference. This improvement is associated with the hierarchical reconstruction,
which progressively recovers fine-scale spatial information through the intermediate
resolution levels and thereby reduces the coarse-grid imprinting observed in the final
prediction compared with the direct $8\rightarrow1$ reconstruction.

At the full-domain scale, however, the difference between the two approaches becomes less
apparent. Figure~\ref{fig:hier_nonhier_global} compares representative two-dimensional
$x-y$ slices at $z=0$ from the unseen test dataset for $\psi_3$ and $\psi_4$. Both
H-NFSR and NFSR recover very similar global flame structures, and the thin coarse-cell
boundaries observed at the local subdomain level become difficult to identify when the
reconstructed field is viewed over the complete fine-grid domain. This shows that the
artifact is primarily a local reconstruction feature rather than a large-scale distortion
of the global field.

The quantitative performance of the two reconstruction strategies remains comparable.
A complete quantitative comparison of all four model configurations is presented in
Section~\ref{sec:model_comparison}. For the unseen test dataset, the mean NMAE over
the 4 PCs is $1.36\%$ for H-NFSR and $1.34\%$ for the
single-stage NFSR, while the corresponding mean SSIM values are 0.9238 and 0.9265,
respectively. Therefore, the two approaches provide nearly identical global reconstruction
accuracy, despite the visible differences observed at the local subdomain level.

These results demonstrate that the single-stage NFSR can achieve comparable, and in some
cases slightly improved, global error metrics on the unseen test dataset. However, the
local reconstructions reveal a difference that is not fully represented by the global
NMAE and SSIM values. The hierarchical approach reduces the visible coarse-cell
imprinting and produces a more spatially continuous reconstruction at the subdomain
level. Accordingly, the main advantage of the hierarchical formulation in the present
problem is associated with the local quality and continuity of the reconstructed field
rather than a substantial reduction in the global reconstruction error.

In terms of computational cost, the two reconstruction strategies also differ substantially.
As reported in Table~\ref{tab:model_training}, the baseline H-NFSR model reaches its
minimum validation loss after 330,000 iterations, corresponding to approximately
6.97~h of training, whereas the single-stage NFSR reaches its best validation point after
112,000 iterations and approximately 0.50~h of training. Hence, the improved local
continuity obtained through the hierarchical reconstruction is accompanied by an increased
training cost. For applications where suppression of local coarse-cell imprinting and
improved spatial continuity are important, the hierarchical formulation provides an
advantage. However, when these local reconstruction characteristics are not a primary
requirement, the single-stage NFSR provides a computationally efficient alternative while
maintaining comparable global reconstruction accuracy.

\subsection{Effect of Bias Sampling in Model Training}
\label{sec:bias_effect}

\begin{figure}[t]
    \centering
    \includegraphics[width=0.92\textwidth]{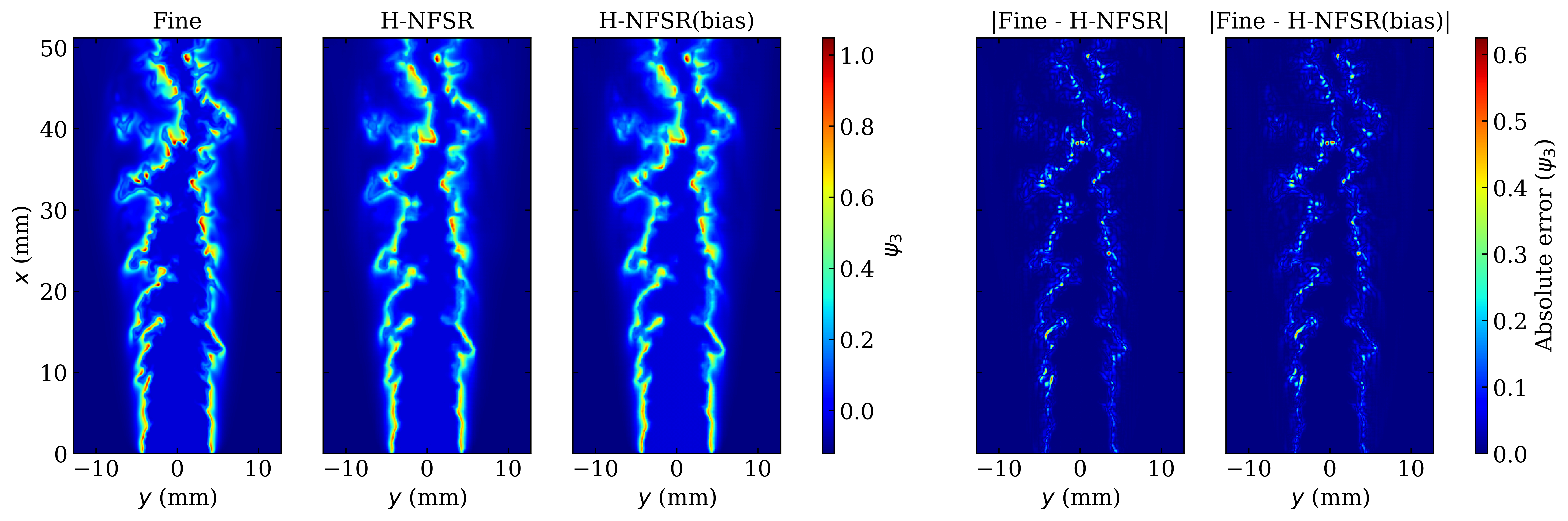}\\[2mm]
    \includegraphics[width=0.92\textwidth]{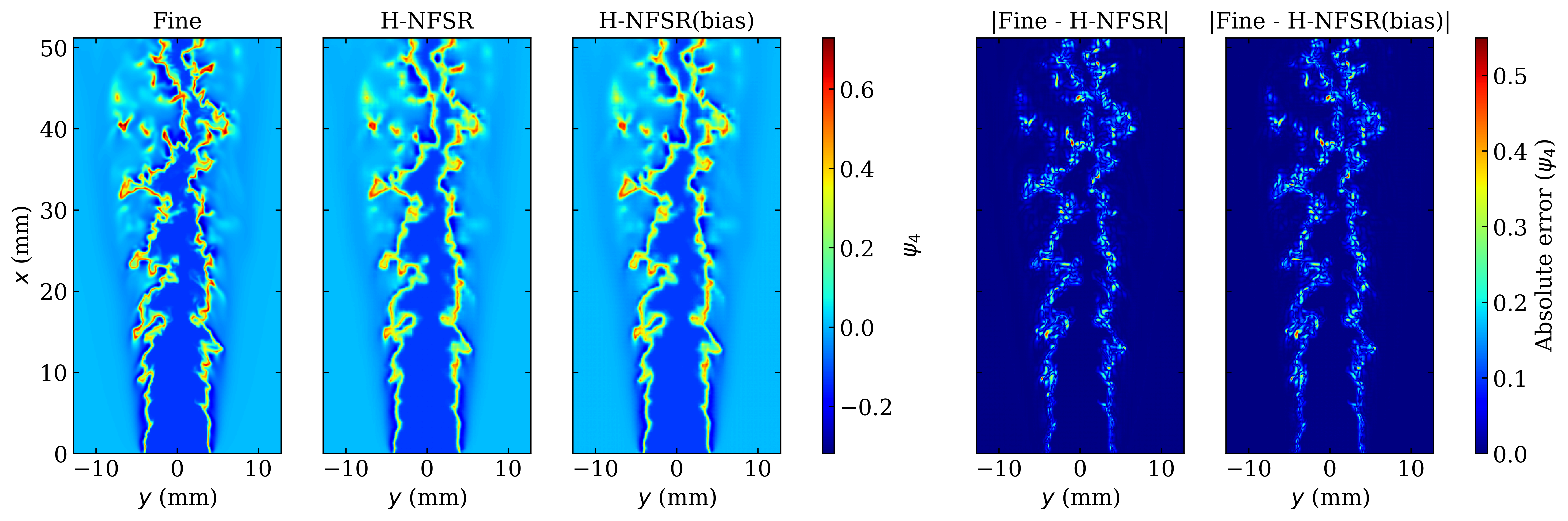}
    \caption{Comparison of the baseline H-NFSR model and H-NFSR with biased sampling
    for representative global $x-y$ slices at $z=0$ from the unseen test dataset.
    From top to bottom, the rows correspond to $\psi_3$ and $\psi_4$, respectively.
    For each PC, the reference fine-grid DNS field, baseline H-NFSR
    reconstruction, biased-sampling H-NFSR reconstruction, and their corresponding
    absolute errors are shown.}
    \label{fig:bias_comparison}
\end{figure}

The effect of the biased sampling strategy is evaluated by comparing the baseline H-NFSR
model with the H-NFSR model trained using biased sampling. The architecture, loss
function, and training procedure remain unchanged between the two cases, such that the
difference is limited to the sampling probability of the local training subdomains. As
described in Section~2.3.2, the biased sampling strategy increases the frequency at which
regions with stronger spatial variations are presented during training.

Figure~\ref{fig:bias_comparison} compares representative $x-y$ slices at $z=0$ from
the unseen test dataset for $\psi_3$ and $\psi_4$. At the global-field level, both models
produce very similar reconstructions and recover the dominant flame-front structures.
Therefore, the effect of biased sampling is considerably less apparent visually than the
difference observed between the hierarchical and single-stage reconstruction strategies.

The quantitative results indicate that the influence of biased sampling is not uniform across
the 4 PCs. A complete comparison of all model configurations is
presented in Section~\ref{sec:model_comparison}. For the unseen test dataset, the mean
NMAE over the 4 PCs changes only from $1.36\%$ for the baseline
H-NFSR model to $1.38\%$ with biased sampling, while the corresponding mean SSIM
increases slightly from 0.9238 to 0.9261. This indicates that biased sampling does not
provide a general reduction in the overall reconstruction error.

A more noticeable improvement is observed for $\psi_4$, which contains the thinnest and
most localized structures among the retained PCs. For this component,
the test NMAE decreases from $1.63\%$ to $1.58\%$, while the SSIM increases from
0.8725 to 0.8854 when biased sampling is introduced. In contrast, the reconstruction of
$\psi_3$ shows little benefit from the biased sampling, with only small changes in both
NMAE and SSIM. These results indicate that increasing the sampling frequency of
high-variation regions can improve the reconstruction of some of the more localized
structures, but does not lead to a consistent improvement across all PCs.

The baseline model nevertheless reconstructs the flame-front structures with good fidelity
without explicitly biasing the training distribution toward high-variation regions. This
suggests that the combination of overlapping subdomains, spatial augmentation, and the
H-NFSR architecture already provides sufficient exposure to the flame-front regions for
the present dataset. The biased sampling strategy therefore provides a localized benefit,
particularly for $\psi_4$, rather than a substantial improvement in the overall
generalization performance.

\subsection{Effect of Loss Function in Super-resolution}
\label{sec:loss_effect}

The effect of the loss function is evaluated by comparing two H-NFSR models trained
using mean squared error (MSE) and mean absolute error (MAE), respectively. All other
model parameters and training settings are maintained unchanged, such that the comparison
isolates the influence of the loss criterion on the reconstruction performance.

\begin{figure}[t]
    \centering
    \includegraphics[width=0.92\textwidth]{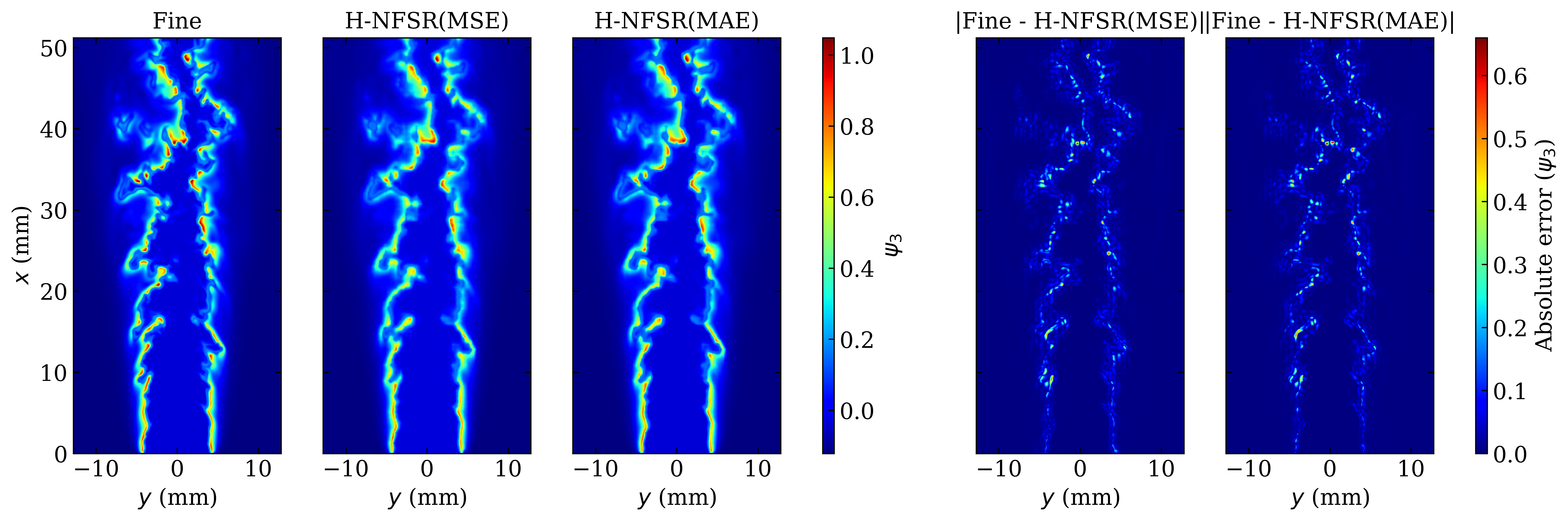}\\[2mm]
    \includegraphics[width=0.92\textwidth]{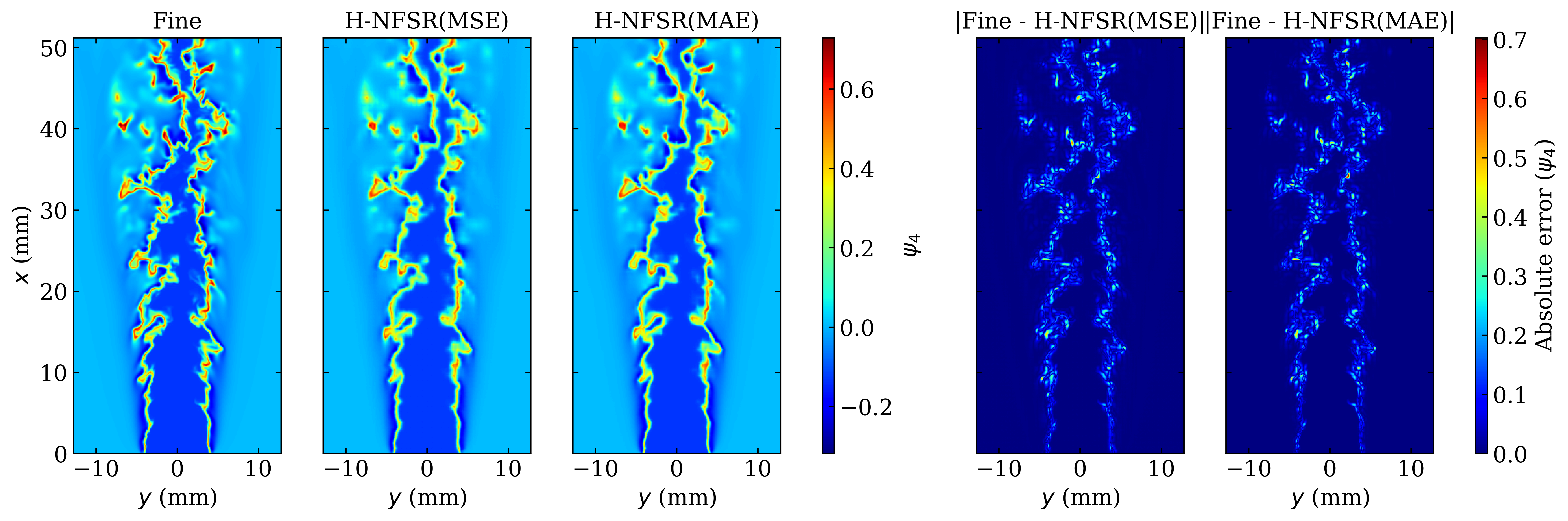}
    \caption{Comparison of the H-NFSR models trained using MSE and MAE for
    representative global $x-y$ slices at $z=0$ from the unseen test dataset.
    From top to bottom, the rows correspond to $\psi_3$ and $\psi_4$, respectively.
    For each PC, the reference fine-grid DNS field, MSE-trained
    H-NFSR reconstruction, MAE-trained H-NFSR reconstruction, and their corresponding
    absolute errors are shown.}
    \label{fig:loss_comparison}
\end{figure}

Figure~\ref{fig:loss_comparison} compares representative $x-y$ slices at $z=0$ from
the unseen test dataset for $\psi_3$ and $\psi_4$. The reconstructions obtained using
MSE and MAE are visually very similar, and both models recover the dominant fine-scale
structures present in the reference DNS field. In particular, the MSE-trained model does
not show a clear increase in smoothing of the thin and localized structures compared with
the MAE-trained model.

In terms of quantitative metrics, MAE training yields a modest improvement.
For the unseen test dataset, the mean NMAE over the 4 PCs decreases
from $1.36\%$ for the MSE-trained model to approximately $1.27\%$ for the
MAE-trained model, while the corresponding mean SSIM increases from 0.9238 to
approximately 0.9294. The improvement is more noticeable for $\psi_4$, where the
NMAE decreases from $1.63\%$ to approximately $1.45\%$, while the SSIM
increases from 0.8725 to approximately 0.8927. A complete quantitative comparison of
all model configurations is presented in Section~\ref{sec:model_comparison}.

Although the MAE-trained model provides slightly improved quantitative performance, the
visual comparison does not indicate a substantial difference in the sharpness of the
reconstructed fields. This behavior differs from the commonly reported tendency of
MSE-trained models to produce smoother reconstructions in some conventional
super-resolution approaches. This behavior may be related to the pointwise formulation used in
the present NFSR framework. During training, queried spatial points are sampled from
different local subdomains within each batch, rather than reconstructing an entire image
or three-dimensional field simultaneously. Therefore, the influence of the loss function
may differ from that observed in architectures that operate directly on complete spatial
patches or volumes. However, the effects of the architecture and sampling procedure are
not independently isolated in the present study, and the observed behavior cannot be
attributed to a single mechanism.

The choice of loss function also influences the convergence behavior. As reported in
Table~\ref{tab:model_training}, the MSE-trained baseline reaches its minimum validation
loss after 330,000 iterations and approximately 6.97~h of training, whereas the
MAE-trained model reaches its best validation point after 564,000 iterations and
approximately 8.62~h. Therefore, the modest improvement obtained using MAE is
accompanied by a longer convergence time. These results indicate that the relative
behavior of MSE and MAE should be interpreted within the context of the model
architecture, sampling procedure, and reconstruction formulation rather than generalized
directly across different super-resolution approaches.

\begin{figure}[t]
\centering

\includegraphics[width=\textwidth]{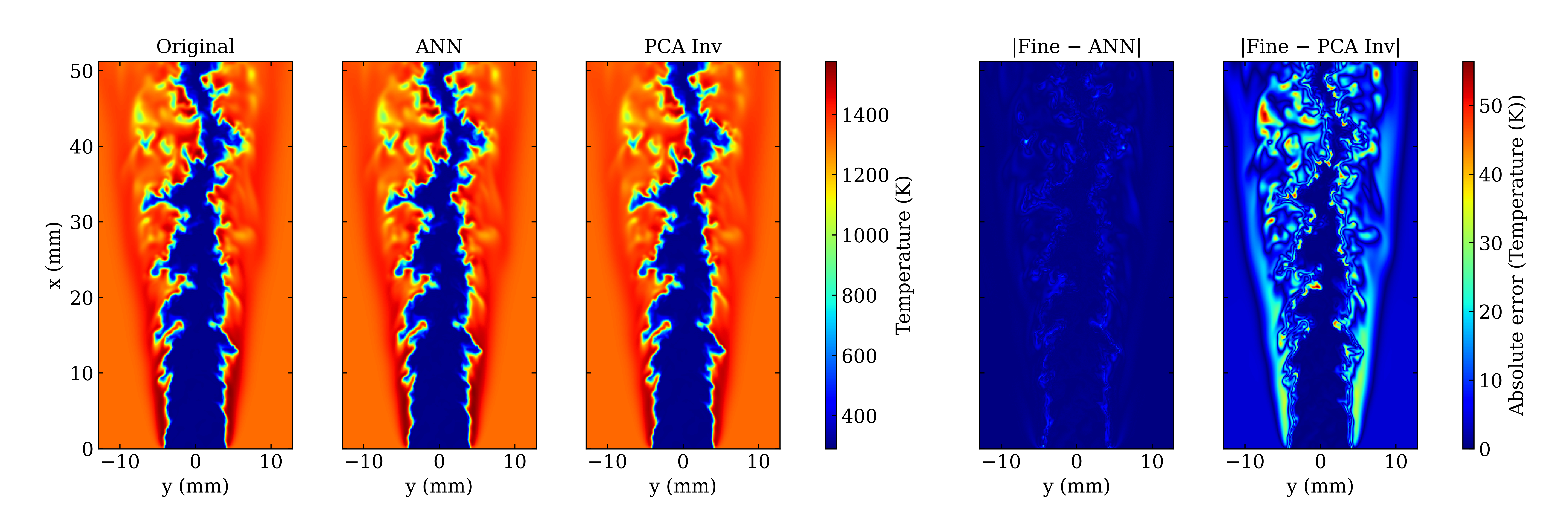}

\includegraphics[width=\textwidth]{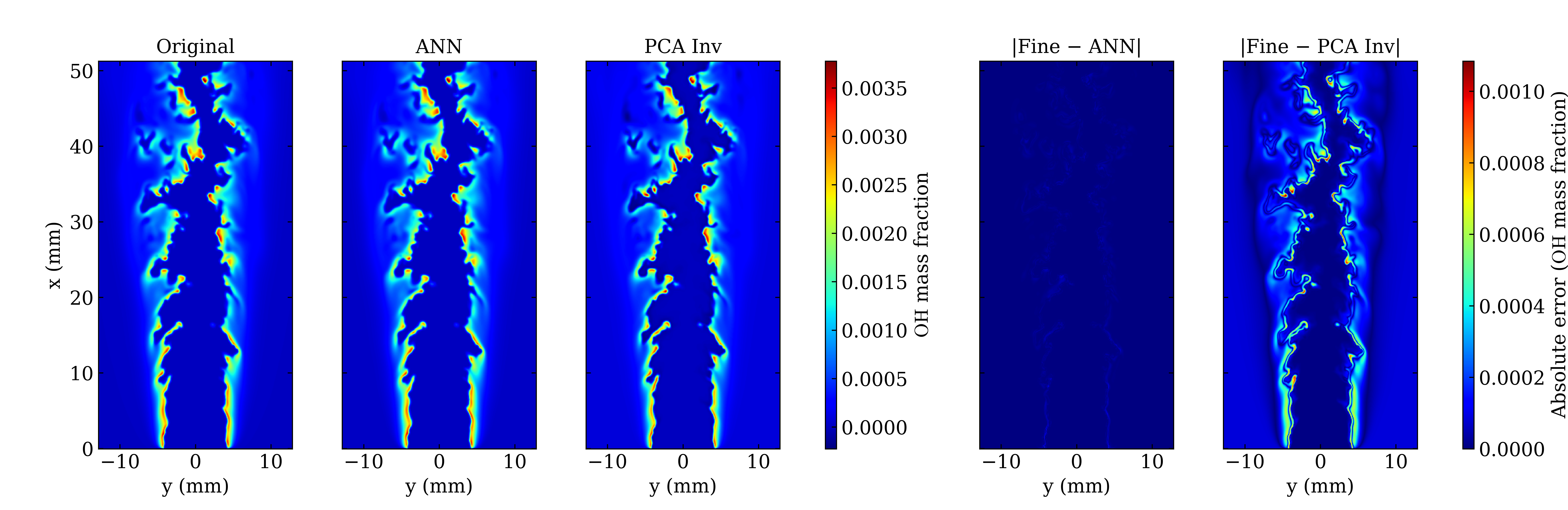}

\caption{Comparison of thermochemical scalar recovery using the nonlinear ANN
mapping and direct inverse PCA for the unseen test realization. From left to right,
the panels show the reference fine-grid field, ANN reconstruction, inverse PCA
reconstruction, absolute error of the ANN reconstruction, and absolute error of the
inverse PCA reconstruction. Temperature and OH mass fraction are shown in the upper
and lower rows, respectively.}
\label{fig:scalar_recovery_ann}

\end{figure}

\begin{figure}[htbp]
\centering

\includegraphics[width=\textwidth]{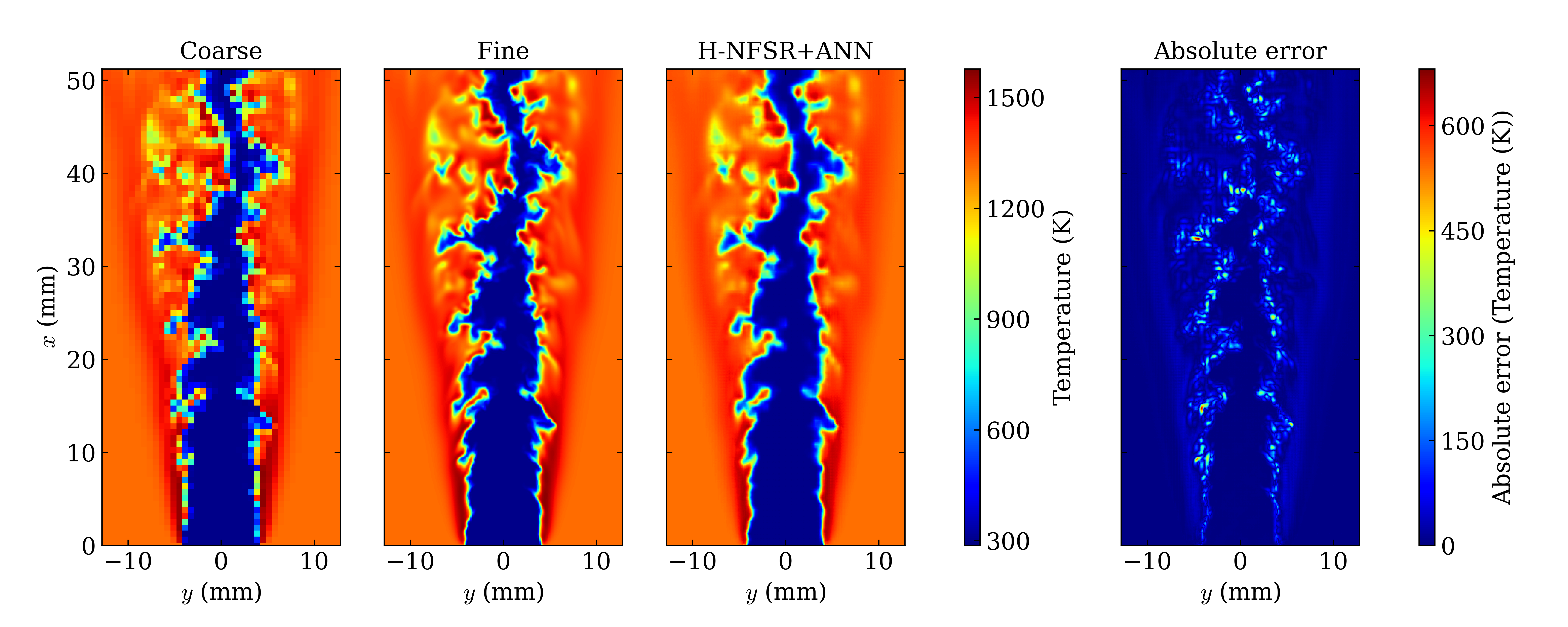}

\includegraphics[width=\textwidth]{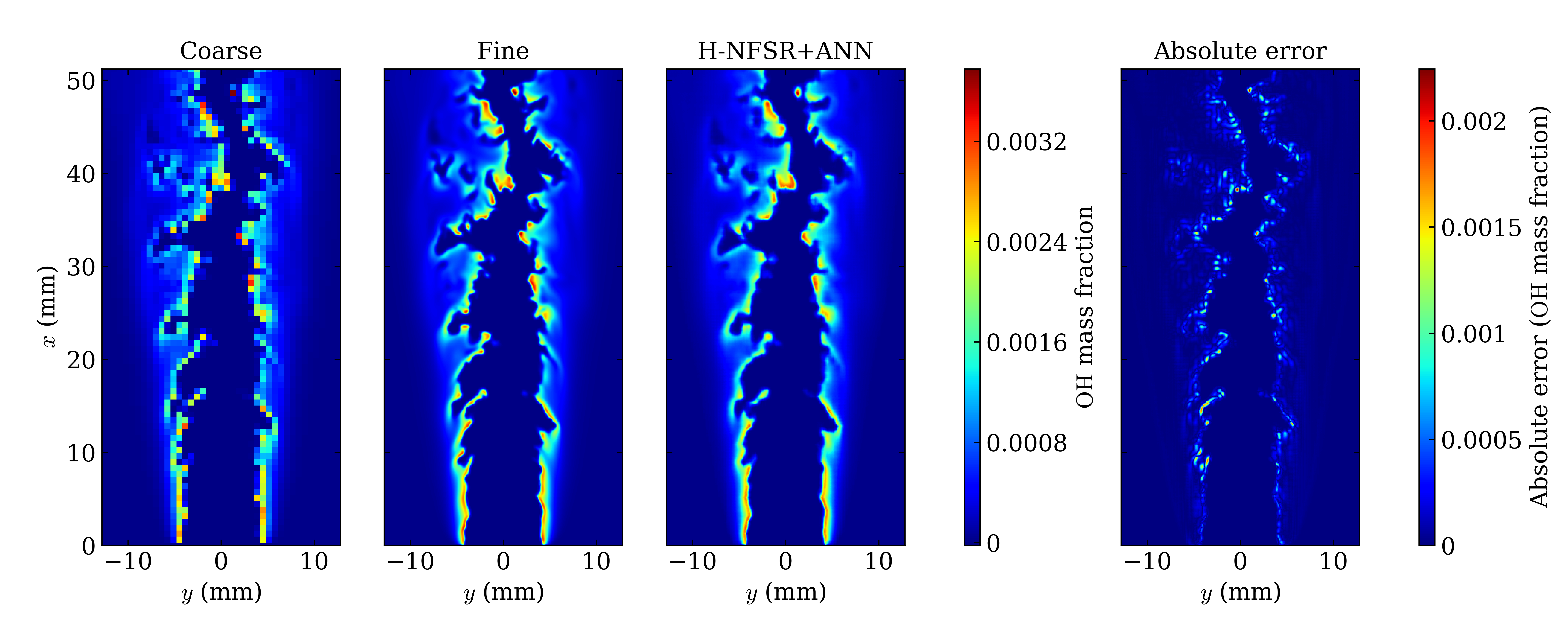}

\caption{End-to-end thermochemical scalar reconstruction using the combined
H-NFSR--ANN framework for the unseen test realization. From left to right, the panels
show the coarse-grid field, reference fine-grid field, H-NFSR--ANN reconstruction,
and corresponding absolute error. Temperature and OH mass fraction are shown in the
upper and lower rows, respectively.}
\label{fig:scalar_recovery_hnfsr}

\end{figure}

\subsection{Thermochemical Scalar Recovery from Retained PCs}
\label{sec:scalar_recovery}

The NFSR framework reconstructs the reacting-flow state in the reduced
principal-component space. To recover the corresponding thermochemical scalar fields
from the four retained PCs, a fully connected artificial neural network
(ANN) is employed. The ANN consists of four hidden layers containing 50, 70, 70, and
50 neurons, respectively. The 4 retained PCs form the network input,
while the outputs correspond to temperature and the eight species mass fractions
introduced in Section~\ref{sec:datadescription}. The ANN is trained using the same
snapshot separation adopted for the NFSR framework, with three snapshots used for
training, one for validation, and the remaining snapshot retained for testing.

This nonlinear mapping through ANN is employed to reduce the reconstruction error associated
with recovering the original thermochemical scalars rather than directly through the inverse PCA
transformation when only a limited number of PCs are retained~\citep{Mirgolbabaei2013, PCA-linear, Mirgolbabaei2015b}.
Figure~\ref{fig:scalar_recovery_ann} compares the ANN reconstruction with direct inverse
PCA for temperature and OH mass fraction using the 4 retained PCs
from the unseen test realization. Temperature is selected to represent the broader
thermochemical field, whereas OH provides a more stringent assessment because of its
thin and highly localized spatial distribution. For both quantities, the ANN reconstruction
closely follows the reference fine-grid field and produces substantially smaller errors than
the direct inverse PCA reconstruction. The improvement is particularly evident in regions
containing strong gradients and localized reaction-zone structures, where the inverse PCA
reconstruction exhibits larger deviations from the reference field.

The improved scalar recovery is also observed across the complete set of reconstructed
thermochemical quantities. The ANN provides a lower normalized mean absolute error
than direct inverse PCA for all nine reconstructed scalars. Averaged over temperature
and the eight species mass fractions, the normalized mean absolute error is reduced by
approximately 94\%. The normalized maximum absolute error is also reduced for eight
of the nine reconstructed quantities. These results support the use of the nonlinear ANN
mapping for recovering the thermochemical state from the reduced principal-component
representation.

The complete coarse-to-fine reconstruction is subsequently evaluated by coupling the
H-NFSR prediction with a trained ANN that maps the reconstructed PC fields back to
the physical thermochemical variables. The ANN takes the four retained PCs as inputs
and predicts nine thermochemical scalars, consisting of temperature and the eight species
mass fractions considered in the present study. The network consists of four fully
connected hidden layers, each containing 70 neurons. This mapping enables the
fine-resolution information reconstructed by H-NFSR in the reduced PC space to be
transferred to the corresponding physical thermochemical fields. 

Figure~\ref{fig:scalar_recovery_hnfsr} shows the resulting end-to-end reconstruction
for temperature and OH mass fraction from the unseen test realization. The figure compares
the coarse-grid field, reference fine-grid field, combined H-NFSR--ANN reconstruction,
and corresponding absolute error. The combined framework recovers much of the
fine-scale thermochemical structure that is absent or poorly represented in the coarse-grid
fields. In particular, the reconstructed flame front geometry, localized wrinkling, and
thin OH structures closely follow the fine-grid reference, while the remaining errors are
concentrated primarily around regions of strong spatial gradients. This comparison shows
that the fine-scale information reconstructed by H-NFSR in the reduced
principal-component space can be transferred effectively to the physical thermochemical
variables through the nonlinear ANN mapping.

\begin{figure}[t]
    \centering

    \begin{minipage}[t]{0.32\textwidth}
        \centering
        \includegraphics[
            trim=600 0 550 0,
            clip,
            height=0.34\textheight,
            keepaspectratio
        ]{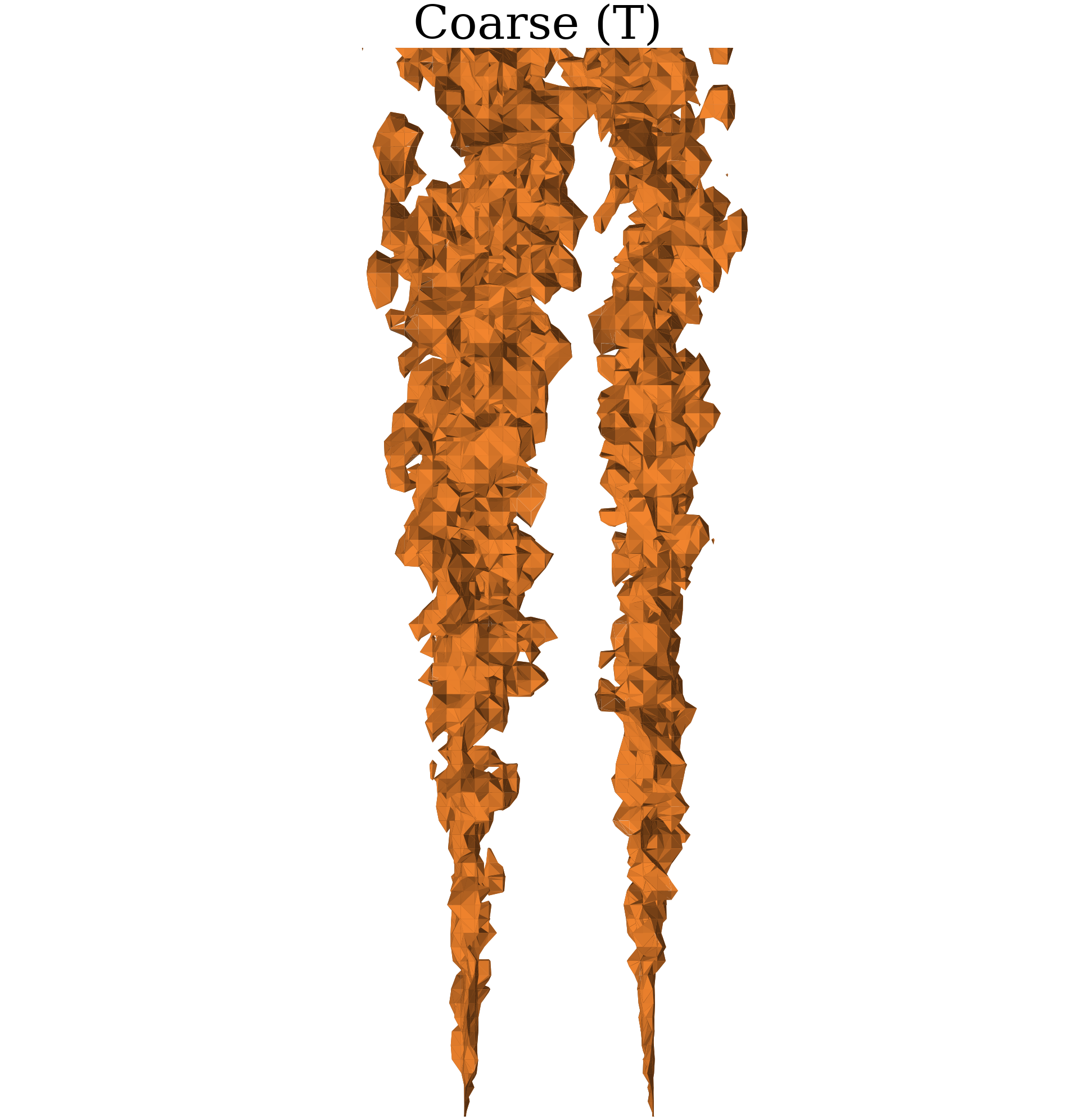}
    \end{minipage}
    \hfill
    \begin{minipage}[t]{0.32\textwidth}
        \centering
        \includegraphics[
            trim=600 0 575 0,
            clip,
            height=0.34\textheight,
            keepaspectratio
        ]{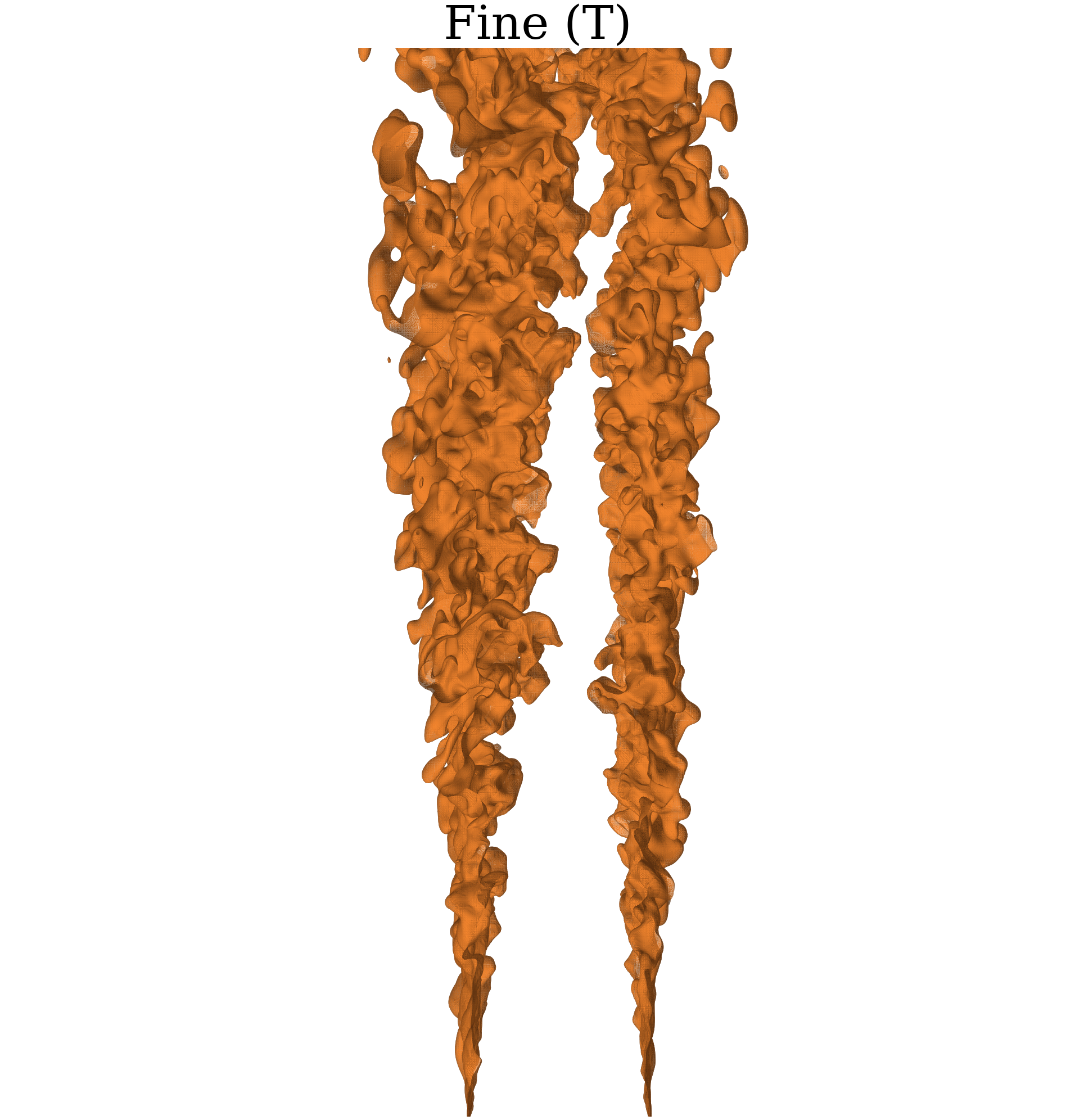}
    \end{minipage}
    \hfill
    \begin{minipage}[t]{0.32\textwidth}
        \centering
        \includegraphics[
            trim=600 0 600 0,
            clip,
            height=0.34\textheight,
            keepaspectratio
        ]{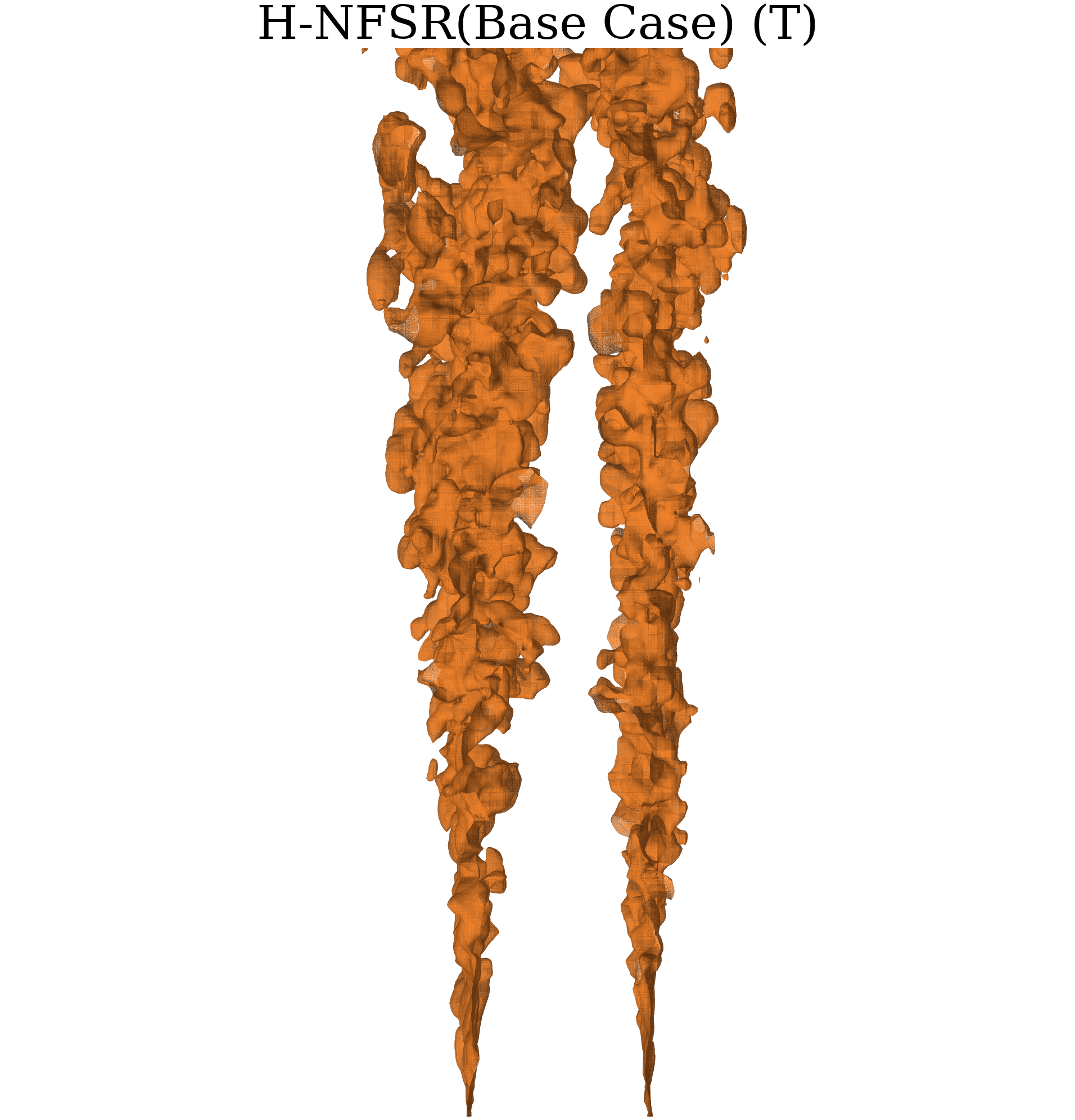}
    \end{minipage}

    \caption{Three-dimensional temperature-isosurface comparison for the unseen test
    realization at $T=1200~\mathrm{K}$. From left to right, the panels show the coarse-grid
    temperature field, reference fine-grid DNS temperature field, and the temperature field
    reconstructed using the combined H-NFSR--ANN framework.}
    \label{fig:T_3d_comparison}
\end{figure}

\subsection{Three-Dimensional Reconstruction of Fine-Scale Structures}
\label{sec:3d_reconstruction}

The preceding comparisons primarily examined local subdomains and two-dimensional
slices of the reconstructed fields. Following the recovery of the physical thermochemical
variables described in Section~\ref{sec:scalar_recovery}, the three-dimensional
reconstruction capability of the complete H-NFSR--ANN framework is further assessed
using the temperature field. Figure~\ref{fig:T_3d_comparison} compares the
$T=1200~\mathrm{K}$ isosurface obtained from the coarse-grid field, the reference
fine-grid DNS field, and the H-NFSR--ANN reconstruction for the unseen test
realization. Temperature is selected for this visualization because it provides a clear and
physically interpretable representation of the overall flame structure and its
three-dimensional geometric complexity. Within the approximate temperature range of
$600$--$1650~\mathrm{K}$ present in the domain, the $1200~\mathrm{K}$ isosurface lies
within the intermediate-temperature flame-transition region and therefore provides a
representative surface for examining the three-dimensional flame morphology.

The coarse-grid representation preserves the overall large-scale scalar structures, but the surface is strongly constrained by the reduced spatial
resolution. The resulting isosurface exhibits a visibly faceted geometry and does not
fully represent many of the smaller-scale variations present in the fine-grid DNS field.
In particular, localized protrusions, surface wrinkling, and several smaller disconnected
structures are either substantially simplified or absent from the coarse representation.

In contrast, the H-NFSR--ANN reconstruction recovers a substantially more detailed
isosurface that closely follows the morphology of the fine-grid DNS reference. Much of
the surface wrinkling, local curvature, and irregular geometric structure that is absent
from the coarse-grid field reappears in the reconstructed temperature field. The
reconstruction also recovers several localized protrusions and disconnected structures
at locations consistent with the fine-grid reference, demonstrating that the improvement
is not limited to smoothing or refinement of the coarse surface but includes the recovery
of additional three-dimensional spatial structure.

This three-dimensional comparison complements the preceding slice-based analysis and
shows that the fine-scale information reconstructed by H-NFSR in the reduced
principal-component space remains physically meaningful after transformation back to
the thermochemical variables through the ANN. The agreement between the reconstructed
and reference temperature isosurfaces further demonstrates that the proposed framework
can recover a substantial portion of the three-dimensional geometric complexity lost
during spatial coarsening. This result is particularly encouraging given that the model is
trained using only three statistically stationary snapshots, which provide limited coverage
of the full range of spatial structures that may occur within the selected operating
condition.

\subsection{Summary of Model Comparisons}
\label{sec:model_comparison}

The preceding sections examined the individual effects of the hierarchical reconstruction,
biased sampling, and loss function on the NFSR framework. A combined quantitative
comparison of these configurations is presented in Table~\ref{tab:overall_model_comparison}
for the unseen test dataset. The comparison includes the baseline H-NFSR model, H-NFSR
with biased sampling, the single-stage NFSR model, H-NFSR trained using MAE, and
tricubic interpolation.

\begin{table}[t]
\centering
\caption{Overall comparison of the NFSR model configurations and tricubic interpolation
for the unseen test dataset. NMAE and MaxAbs are reported as percentages.}
\label{tab:overall_model_comparison}

\footnotesize
\setlength{\tabcolsep}{4pt}
\renewcommand{\arraystretch}{1.15}

\begin{tabular}{llccccc}
\hline
\noalign{\vskip 4pt}
Metric & PC &
\shortstack{H-NFSR\\no bias} &
\shortstack{H-NFSR\\bias} &
\shortstack{NFSR\\no bias} &
\shortstack{H-NFSR\\MAE} &
Tricubic \\
\hline

NMAE (\%)
& $\psi_1$ & 1.50 & 1.60 & 1.54 & \textbf{1.42} & 1.76 \\
& $\psi_2$ & 1.36 & 1.37 & 1.36 & \textbf{1.31} & 1.71 \\
& $\psi_3$ & 0.95 & 0.96 & 0.93 & \textbf{0.90} & 1.36 \\
& $\psi_4$ & 1.63 & 1.58 & 1.55 & \textbf{1.45} & 2.37 \\
& Mean & 1.36 & 1.38 & 1.34 & \textbf{1.27} & 1.80 \\[1mm]

\hline

MaxAbs (\%)
& $\psi_1$ & 72.09 & 66.12 & 82.39 & 81.85 & \textbf{55.67} \\
& $\psi_2$ & 73.68 & 70.48 & 73.79 & 83.42 & \textbf{57.28} \\
& $\psi_3$ & 65.81 & \textbf{64.28} & 79.51 & 71.85 & 67.15 \\
& $\psi_4$ & 57.86 & 59.57 & \textbf{57.69} & 61.62 & 67.12 \\[1mm]

\hline

SSIM
& $\psi_1$ & 0.9521 & 0.9511 & 0.9512 & \textbf{0.9529} & 0.9297 \\
& $\psi_2$ & 0.9324 & 0.9324 & 0.9325 & \textbf{0.9332} & 0.9033 \\
& $\psi_3$ & 0.9383 & 0.9354 & 0.9385 & \textbf{0.9386} & 0.8928 \\
& $\psi_4$ & 0.8725 & 0.8854 & 0.8838 & \textbf{0.8927} & 0.7997 \\
& Mean & 0.9238 & 0.9261 & 0.9265 & \textbf{0.9294} & 0.8814 \\
\hline
\end{tabular}
\end{table}

The learned reconstruction models provide broadly comparable global performance, while
all four NFSR configurations substantially improve the NMAE and SSIM relative in comparison with
tricubic interpolation. The differences among the learned models are considerably smaller
than the differences between the learned reconstructions and tricubic interpolation,
indicating that the NFSR formulation itself provides the primary improvement in the
recovery of the fine-grid fields.

Among the learned configurations, the MAE-trained H-NFSR model provides the lowest
mean NMAE over the 4 PCs, approximately $1.27\%$, together
with the highest mean SSIM of approximately 0.9294. The single-stage NFSR also
provides competitive global performance, with a mean NMAE of approximately
$1.34\%$ and mean SSIM of 0.9265, compared with $1.36\%$ and 0.9238,
respectively, for the baseline H-NFSR model. However, as demonstrated in
Section~\ref{sec:hierarchical_effect}, these global metrics do not fully represent the
improved local spatial continuity obtained through the hierarchical reconstruction.

The effect of biased sampling is similarly dependent on the reconstructed principal
component. Its overall quantitative performance remains close to that of the baseline
model, while a more noticeable improvement is obtained for $\psi_4$, where the
localized spatial structures are more strongly represented during training. Likewise,
the MAE-trained model provides modest improvements in the global NMAE and SSIM,
particularly for $\psi_4$, although these improvements are accompanied by a longer
training time.

The MaxAbs metric exhibits a less systematic trend across the different reconstruction
approaches. In particular, a lower maximum error does not necessarily correspond to
improved reconstruction over the majority of the spatial domain, as illustrated by the
tricubic results for $\psi_1$ and $\psi_2$. The NMAE and SSIM therefore provide a more
representative measure of the overall and structural reconstruction quality, while MaxAbs
identifies the largest isolated pointwise discrepancies.

Overall, the comparisons indicate that the MAE-trained H-NFSR provides the
strongest overall quantitative performance among the learned configurations, particularly
in terms of NMAE and SSIM. However, the differences among the NFSR variants remain
relatively modest compared with the larger improvement obtained over tricubic
interpolation. The hierarchical formulation primarily improves local spatial continuity,
while biased sampling provides a more localized benefit for strongly varying structures,
particularly for $\psi_4$. The single-stage NFSR achieves comparable global accuracy
with a substantially lower training cost. These results therefore show that the different
model configurations provide similar overall reconstruction capability, with the choice
among them depending on the desired balance between quantitative accuracy, local
continuity, and computational cost.


\section{Conclusions}
\label{sec:Conclusion}

This study developed a NeRF-inspired Neural Field Super-Resolution (NFSR) framework
for reconstructing three-dimensional fine-resolution reacting-flow fields from spatially
coarsened solution data. The proposed formulation replaces the view-dependent radiance
prediction of the original NeRF architecture with a coordinate-conditioned mapping that
combines local spatial information with coarse-grid PC fields. A hierarchical extension,
H-NFSR, performs the reconstruction progressively through intermediate resolution
levels, while PCA provides a compact and low-dimensional representation of the thermochemical state. The
framework was evaluated using three-dimensional DNS data for a lean premixed
hydrogen--air flame with an overall spatial coarsening factor of eight in each coordinate
direction.

The results demonstrate that the proposed neural-field formulation can recover
fine-scale spatial information that is not directly represented in the coarse-grid input.
Compared with tricubic interpolation, the learned reconstructions provide improved
overall accuracy and structural similarity across the retained PC fields. The reconstructed
solutions recover thin interfaces, localized structures, and small-scale spatial variations
that are strongly attenuated in the coarse representation. These improvements are
particularly evident for the higher-order PCs, which contain thinner and more localized
spatial structures.

The reconstruction in the reduced PC space is subsequently mapped back to the original
thermochemical variables using an ANN trained to recover temperature and the
eight retained species mass fractions from the four PCs. Compared with direct inverse
PCA, the ANN provides substantially lower reconstruction errors across the recovered
thermochemical quantities. The combined H-NFSR--ANN framework also recovers much
of the fine-scale structure of the physical temperature and OH fields for the unseen test
realization. This demonstrates that the spatial information recovered in the reduced
representation can be transferred effectively back to physically interpretable
thermochemical quantities.

The model comparisons also clarify the influence of the different components of the
framework. The hierarchical reconstruction improves local spatial continuity and reduces
visible coarse-cell imprinting, while the single-stage NFSR provides comparable global
reconstruction accuracy at a substantially lower training cost. Biased sampling provides
a more localized improvement for strongly varying structures, particularly for the
higher-order PCs. Training with MAE provides modest improvements in the overall
quantitative performance compared with MSE, although the reconstructed fields remain
visually similar and the MAE-trained model requires a longer convergence time. These
results show that the NFSR framework can be adapted according to the desired balance
between reconstruction accuracy, local spatial quality, and computational cost.

The three-dimensional reconstruction further demonstrates that the capability of the
framework extends beyond individual planar sections. The temperature field recovered
using the combined H-NFSR--ANN framework reproduces much of the surface wrinkling,
local curvature, and localized three-dimensional structures observed in the fine-grid DNS
reference but poorly represented in the coarse-grid field. This provides a direct
demonstration that the proposed coarse-to-fine reconstruction can recover physically
meaningful three-dimensional thermochemical structure after transformation from the
reduced PC representation.

Despite being trained using a limited number of statistically stationary snapshots, the
framework successfully reconstructs the principal fine-scale and three-dimensional
features of an unseen test realization. This provides encouraging evidence that the
coordinate-conditioned formulation can learn spatial reconstruction relationships that
extend beyond the individual structures presented during training. However, the limited
number of available snapshots does not provide sufficient coverage of the full statistical
variability of the selected operating condition. Future work will therefore extend the
framework to a broader range of snapshots, operating conditions, and reacting-flow
configurations, and will further evaluate its application to independently generated
coarse-grid simulations.

Overall, the present results demonstrate that coordinate-conditioned neural fields provide
an effective approach for three-dimensional super-resolution of reacting-flow fields. The
proposed NFSR framework, together with the nonlinear recovery of physical
thermochemical variables, provides a flexible coarse-to-fine reconstruction strategy for
recovering fine-scale reacting-flow structure from strongly coarsened representations and
establishes a foundation for future application to high-fidelity reacting-flow simulations.

\bibliographystyle{plainnat}
\bibliography{references}

\end{document}